\pdfoutput=1
\documentclass[aip,jcp,reprint,amsmath,amssymb]{revtex4-2}
\usepackage[utf8]{inputenc}
\usepackage[T1]{fontenc}
\usepackage{graphicx}
\usepackage{xurl}
\usepackage{alltt}
\usepackage{booktabs}
\usepackage{multirow}
\usepackage{float}
\usepackage{needspace}
\usepackage[explicit]{titlesec}
\usepackage{fancyhdr}
\usepackage{hyperref}
\hypersetup{colorlinks=true,allcolors=blue}
\fancypagestyle{plain}{\fancyhf{}\cfoot{\thepage}}
\renewcommand{\thesection}{\arabic{section}}
\renewcommand{\thesubsection}{\thesection.\arabic{subsection}}
\renewcommand{\thesubsubsection}{\thesubsection.\arabic{subsubsection}}
\titleformat{\section}{\normalfont\normalsize\bfseries}{\thesection}{0.6em}{#1}
\titleformat{name=\section,numberless}{\normalfont\normalsize\bfseries}{}{0pt}{#1}
\titleformat{\subsection}{\normalfont\normalsize\bfseries}{\thesubsection}{0.6em}{#1}
\titleformat{\subsubsection}{\normalfont\normalsize\bfseries\itshape}{\thesubsubsection}{0.6em}{#1}
\titlespacing*{\section}{0pt}{1.2ex plus 0.4ex minus 0.2ex}{0.8ex plus 0.2ex}
\titlespacing*{\subsection}{0pt}{1.0ex plus 0.4ex minus 0.2ex}{0.6ex plus 0.2ex}
\titlespacing*{\subsubsection}{0pt}{0.8ex plus 0.4ex minus 0.2ex}{0.5ex plus 0.2ex}

\begin{document}

\onecolumngrid
\begin{center}
{\large\bfseries Application of Pretrained Universal Machine Learning Force-Fields for Phase Transitions Prediction in ZrO\textsubscript{2}}\\[1.2ex]
Rajan Khadka\textsuperscript{1*}, Axel van de Walle\textsuperscript{2}, Patrick Tepesch\textsuperscript{1*}\\
\textit{\textsuperscript{1}}\textit{Corning Research and Development Corporation, Corning, NY 14870, USA}\\
\textit{\textsuperscript{2}}\textit{School of Engineering, Brown University, Box D, Providence, RI, 02912, USA}\\
*Corresponding authors: \href{mailto:khadkar2@Corning.com}{khadkar2@Corning.com}, \href{mailto:TepeschPD@Corning.com}{TepeschPD@Corning.com}
\end{center}
\begin{center}\begin{minipage}{0.92\textwidth}
\noindent Pretrained universal machine-learning force fields can reduce the cost of finite-temperature molecular dynamics relative to density functional theory. However, most benchmarks test only static properties and do not assess whether these models reproduce the correct phase-transition sequence. Here, we benchmark ten open-source universal machine-learning force fields for zirconia (ZrO\textsubscript{2}) using phonon dispersion calculations and variable-cell isothermal-isobaric molecular dynamics. Zirconia is a useful test because it has three thermodynamically stable crystalline phases prior to melting, following the sequence monoclinic$\rightarrow$tetragonal$\rightarrow$cubic. Four models reproduce this sequence: 7net-l3i5, MatterSim-v1.0.0-1M, MACE-mpa, and MACE-D3. Models with orthorhombic--tetragonal energy errors above about 10 meV/atom (7net-0, CHGNet, and MACE) instead produce an orthorhombic intermediate and predict monoclinic$\rightarrow$orthorhombic$\rightarrow$tetragonal$\rightarrow$cubic$\rightarrow$liquid upon heating. ORB-based models fail to sustain stable variable-cell dynamics because their phonon frequencies are compressed by 30--60\% relative to the density functional theory reference. Across all ten models, two ground-state diagnostics predict finite-temperature behavior. Relative polymorph energetics should agree with the matched density functional theory reference within about 6 meV/atom, and the phonon spectrum should not show systematic softening. These diagnostics are practical screening metrics for assessing pretrained force fields for phase-transition predictions.

\noindent\textbf{Keywords:} machine-learning interatomic potentials, zirconia, molecular dynamics, phase transitions, phonon dispersion, polymorph energetics, dispersion correction, structural phase transitions.
\end{minipage}\end{center}
\twocolumngrid

\section{INTRODUCTION}

Machine-learning interatomic potentials (MLIPs) enable virtual materials screening and design with atomic scale simulations at nanometer and nanosecond scales, beyond the reach of classical force fields and ab initio molecular dynamics (AIMD) [1]. MLIPs map the local atomic environments onto high-dimensional descriptors, allowing the training of force fields that reproduce first-principles energies, forces, and stresses with small errors. Within this broad class, pretrained universal machine-learning force fields (uMLFFs) have emerged as versatile tools. These models are trained on large, heterogeneous datasets and are designed to be transferable across chemical spaces spanning up to $\sim$90 elements of the periodic table. For crystalline materials, numerous uMLFFs now use different architectures and training sets. Most uMLFFs are trained on 0 K density functional theory (DFT) relaxations from the Materials Project [2] (MPtraj) and/or on curated datasets that include AIMD trajectories and relaxed structures. Public benchmarks such as MatBench Discovery [3] report global regression and classification metrics for these models. However, whether a uMLFF that performs well on static benchmarks will also correctly reproduce finite-temperature phase-transition pathways in molecular dynamics (MD) simulations has not been systematically tested, and no diagnostic has been established to predict this from ground-state properties alone.

Studies in 2024--2025 have begun to benchmark multiple aspects of uMLFF accuracy and reliability. MatBench Discovery [3] provides a framework for assessing energy models at scale. Wines et al. [4] developed the CHIPS-FF infrastructure for evaluating open-source uMLFFs on structure optimization and ground-state properties, including elastic constants, phonon dispersion, surface energies, vacancy energies, and associated errors. Chiang et al. [5] introduced MLIP Arena to examine chemical reactivity, stability under extreme conditions, and the prediction of thermodynamic and physical properties. Together, these benchmarks show that uMLFFs can reproduce many static properties with low error. Their main practical advantage, however, is the ability to predict finite-temperature dynamical properties such as phase transitions at far lower cost than AIMD. Reliable prediction of a phase-transition sequence places two demands on a uMLFF. First, the relative energetics of competing polymorphs must be reproduced closely enough that the thermodynamically correct phase remains favored. Second, the vibrational spectrum and anharmonic response of each phase must be captured well enough to sustain stable variable-cell dynamics. Whether current pretrained uMLFFs meet both demands, and which ground-state quantities best predict finite-temperature success, remains open. To our knowledge, Ref. [6] is among the few studies that have tested multiple uMLFFs to simulate ferroelectric tetragonal-to-cubic phase transitions in PbTiO\textsubscript{3}.

In this work, we demonstrate metrics that a uMLFF should reproduce to reliably predict finite-temperature phase-transition sequences in MD simulations. We focus on zirconia (ZrO\textsubscript{2}), a technologically important transition-metal oxide used in thermal barrier coatings, solid oxide fuel cells, and biomedical applications that has three stable polymorphs at ambient pressure prior to melting. Upon heating, ZrO\textsubscript{2} undergoes a sequence of structural phase transitions from a low-symmetry monoclinic phase (P2\textsubscript{1}/c, M) to a tetragonal phase (P4\textsubscript{2}/nmc, T), then to a high-symmetry cubic fluorite phase ($Fm\bar{3}m$, C), before melting (L) [7] (see Figure 1).

Using ZrO\textsubscript{2} as a test case, we evaluate a set of uMLFFs with diverse architectures and training sets, including CHGNet [2], MatterSim [8], SevenNet [9], MACE [10], and ORB [11]. We first benchmark ground-state properties of ZrO\textsubscript{2} phases, including lattice parameters, equations of state (EOS), bulk moduli, and phonon dispersions, against DFT results in Section 2.1. We then perform finite-temperature isothermal-isobaric (NPT) MD simulations at zero pressure using these uMLFFs to directly probe the phase-transition sequence and compare the results with both ``on-the-fly'' VASPML-SVD [7] and experimental data in Section 2.2. Details of the calculations are presented in Section 5. We then correlate the MD-predicted phase behavior with the ground-state benchmarks to identify two practical diagnostics---relative polymorph energy error against DFT and phonon-dispersion fidelity---that predict phase-transition fidelity in Section 3.4. Section 4 summarizes the resulting pre-flight diagnostic for applying pretrained uMLFFs to phase-transition prediction.

\begin{figure*}[tbp]
\centering
\includegraphics[width=\textwidth]{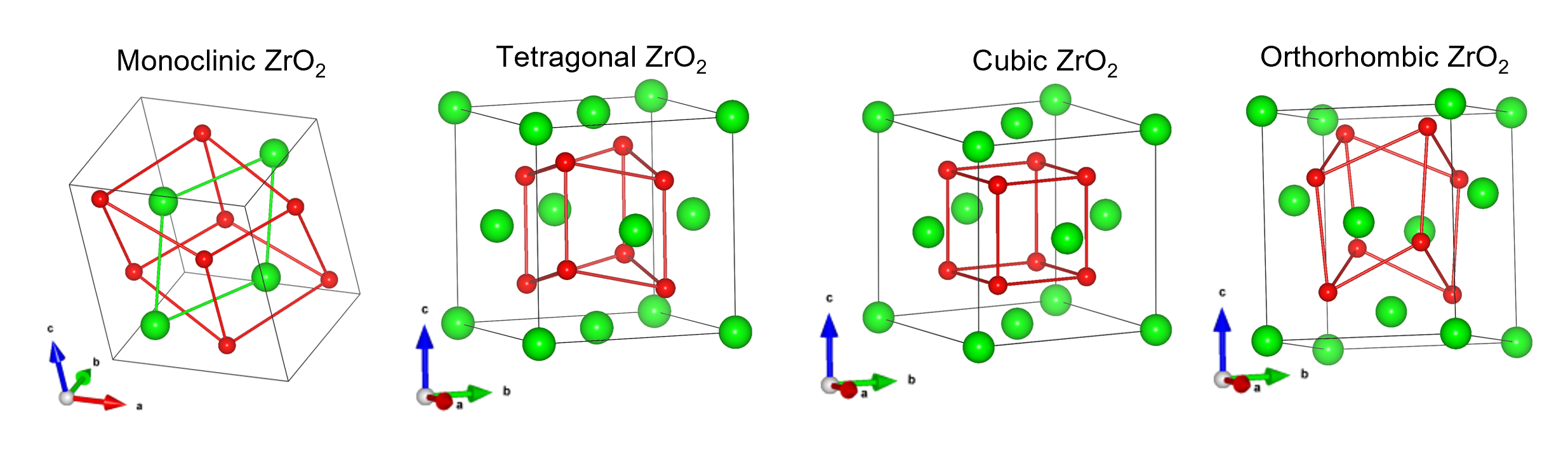}
\caption{\textbf{Crystal structures of the ZrO}\textbf{\textsubscript{2}} \textbf{polymorphs considered in this study.} From left to right: monoclinic (\textit{P}2\textsubscript{1}/\textit{c}, M), tetragonal (\textit{P}4\textsubscript{2}/\textit{nmc}, T), cubic (\textit{Fm}$\bar{3}$\textit{m}, C), and orthorhombic (\textit{Pbcn}, O) ZrO\textsubscript{2}. The monoclinic, tetragonal, and cubic phases form the experimentally established ambient-pressure heating sequence, whereas the orthorhombic phase is included as the metastable intermediate observed in several uMLFF-driven MD simulations. Green and red spheres denote Zr and O atoms, respectively.}
\label{fig:1}
\end{figure*}

\section{RESULTS}

\subsection{Ground-state properties}

In this section, we present lattice parameters, bulk moduli, equations of state (EOS), and phonon dispersions for the monoclinic (M), tetragonal (T), cubic (C), and orthorhombic (O) polymorphs of ZrO\textsubscript{2}. In addition to the M, T, and C phases, we characterize the O phase because it was observed during our phase-transition MD simulations (Section 2.2.2). We note that the O phase is not part of the experimentally established ambient-pressure phase sequence of ZrO\textsubscript{2}; however, a corresponding structure exists in the NIST JARVIS database (JVASP-151841), and its role as a metastable competitor is analyzed in Section 3.3.

\subsubsection{Lattice parameters}

\begin{figure*}[tbp]
\centering
\includegraphics[width=\textwidth,height=0.85\textheight,keepaspectratio]{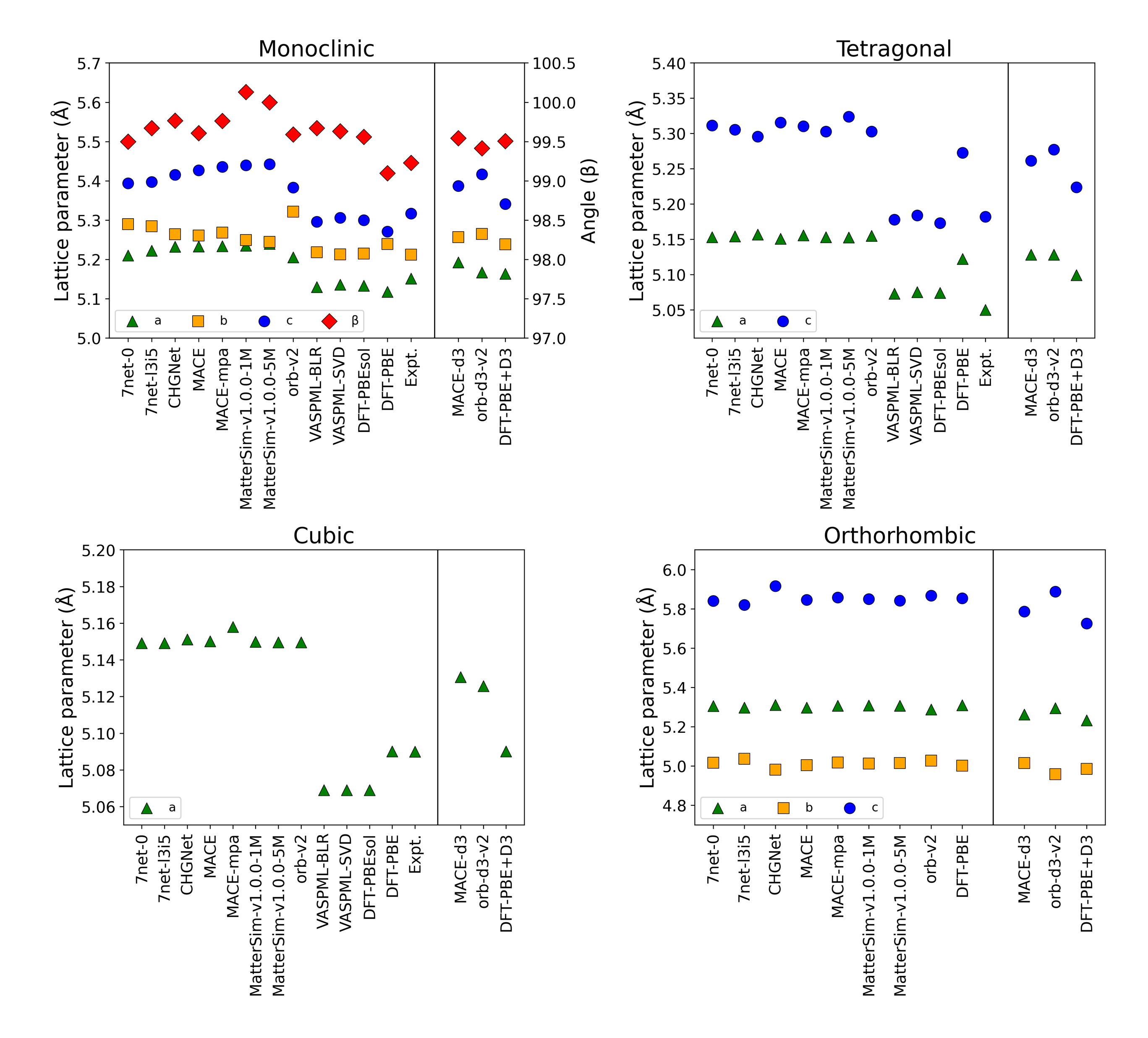}
\caption{\textbf{Lattice parameter benchmarks for ZrO}\textbf{\textsubscript{2}} \textbf{polymorphs at 0 K.}  Lattice parameters of the monoclinic (M), tetragonal (T), cubic (C), and orthorhombic (O) phases predicted by pretrained uMLFFs and compared against DFT references. For the monoclinic panel, the monoclinic angle $\beta$ is shown on a secondary axis.}
\label{fig:2}
\end{figure*}

Figure 2 compares the lattice parameters obtained from ten pretrained uMLFFs with DFT-PBE, DFT-PBE+D3 (this work), and DFT-PBEsol results [7]. Across all phases, the non-dispersion-corrected (non-DC; here ``dispersion'' refers to van der Waals interactions) uMLFFs predict lattice constants that agree with one another to within $\sim$0.2--0.7\% but systematically overestimate DFT-PBE values. For the cubic phase, non-DC uMLFFs show 1.2--1.3\% overestimation with respect to DFT-PBE. For the lower-symmetry monoclinic phase, deviations are larger and parameter-dependent: +1.7 to +2.4\% for a, +0.1 to +1.6\% for b, and +2.1 to +3.3\% for c. PBE and PBEsol themselves also yield different lattice parameters, reflecting the underlying exchange--correlation functional used to generate the training data. Li et al. [6] also reported overestimation of lattice parameters by uMLFFs (CHGNet 0.3.0 and MACE) trained on PBE-based databases in the case of PbTiO\textsubscript{3}. For both MACE-D3 versus MACE and ORB-D3-v2 versus ORB-v2, the dispersion-corrected variants yield smaller lattice parameters across all ZrO\textsubscript{2} phases. When compared against DFT-PBE+D3 $-$ the matched functional for these models $-$ MACE-D3 and ORB-D3-v2 show excellent agreement, with deviations comparable in magnitude to those of the non-DC uMLFFs against plain DFT-PBE.

\subsubsection{Equation of state (EOS)}

\begin{figure*}[tbp]
\centering
\includegraphics[width=\textwidth,height=0.85\textheight,keepaspectratio]{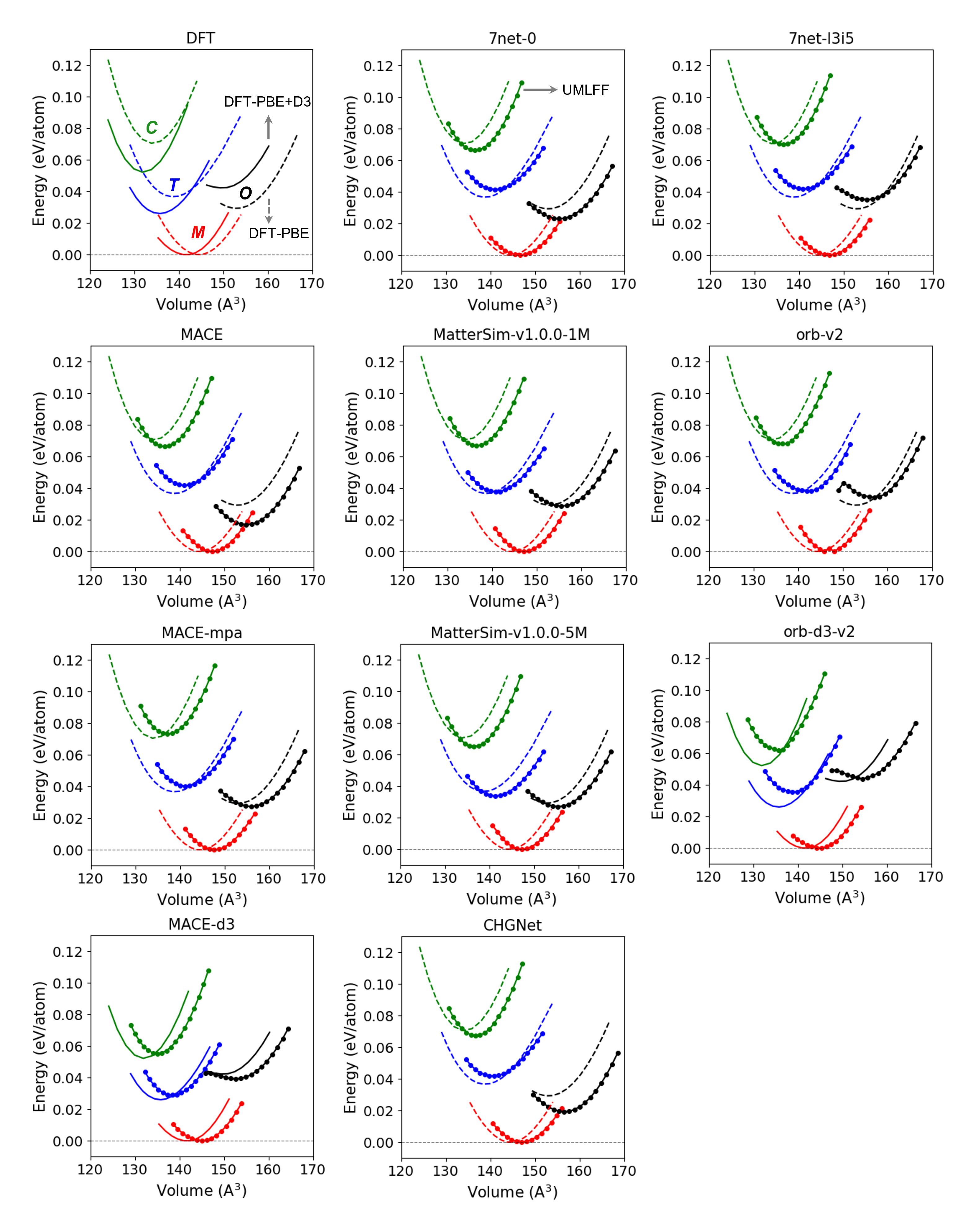}
\caption{\textbf{Equations of state (at 0 K) for the monoclinic (M), orthorhombic (O), tetragonal (T), and cubic (C) phases of ZrO\textsubscript{2}. Energy--volume (E--V) curves are shown for each uMLFF alongside the corresponding DFT reference curves.} The DFT subpanel shows both DFT-PBE (dashed) and DFT-PBE+D3 (solid) curves for direct comparison; subsequent subpanels compare each uMLFF against its matched DFT reference $-$ DFT-PBE for non-dispersion-corrected models and DFT-PBE+D3 for MACE-D3 and ORB-D3-v2. All energies are shifted by the monoclinic ground-state energy E\textsubscript{M} to highlight relative phase stabilities.}
\label{fig:3}
\end{figure*}

We next computed the equations of state for all ZrO\textsubscript{2} phases. Figure 3 shows the energy--volume (E--V) curves obtained from DFT and from the uMLFFs for the M, O, T, and C phases. For direct comparison, all energies---both DFT and uMLFF---were shifted by subtracting the ground-state energy of the monoclinic phase. We observe that the DFT-PBE ground-state energy ordering is \textit{E}\textit{\textsubscript{M}}\textit{<E}\textit{\textsubscript{O}}\textit{<E}\textit{\textsubscript{T}}\textit{<E}\textit{\textsubscript{C}}. All uMLFFs reproduce this ordering except the dispersion-corrected models MACE-D3 and ORB-D3-v2, which instead yield \textit{E}\textit{\textsubscript{M}}\textit{<E}\textit{\textsubscript{T}}\textit{<E}\textit{\textsubscript{O}}\textit{<E}\textit{\textsubscript{C}}. Our DFT-PBE+D3 reference calculations (shown in the DFT panel of Figure 3) confirm that this inverted O/T ordering is an intrinsic feature of the D3-corrected functional itself: DFT-PBE+D3 yields E\textsubscript{T}<E\textsubscript{O}, the same ordering as MACE-D3 and ORB-D3-v2.

\begin{figure*}[tbp]
\centering
\includegraphics[width=\textwidth,height=0.85\textheight,keepaspectratio]{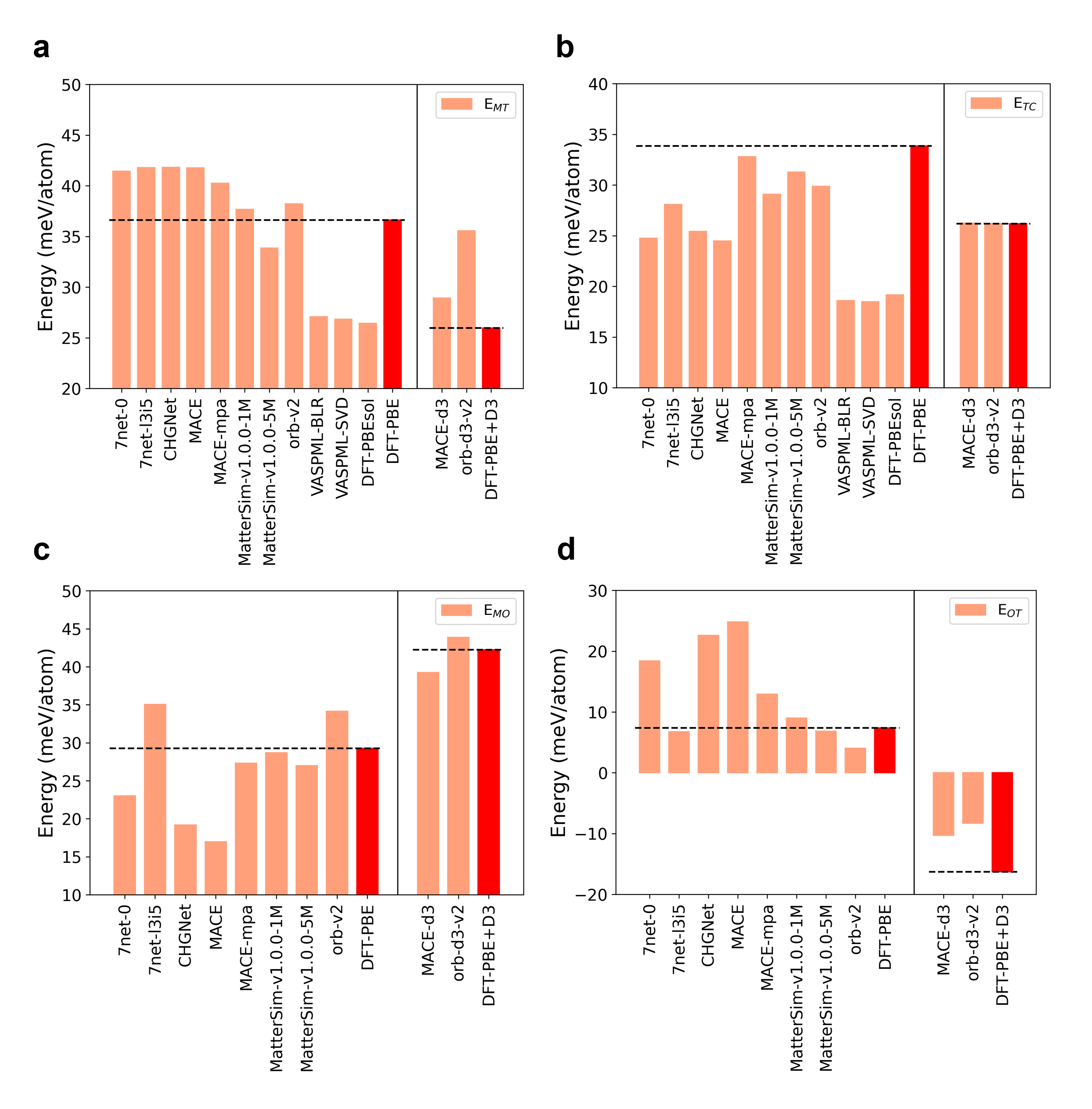}
\caption{\textbf{Relative ground-state energy differences between ZrO}\textbf{\textsubscript{2}} \textbf{polymorphs predicted by uMLFFs and DFT.} (a) Monoclinic--tetragonal, \textit{E}\textsubscript{MT} = \textit{E}\textsubscript{T} $-$ \textit{E}\textsubscript{M}; (b) tetragonal--cubic, \textit{E}\textsubscript{TC} = \textit{E}\textsubscript{C} $-$ \textit{E}\textsubscript{T}; (c) monoclinic--orthorhombic, \textit{E}\textsubscript{MO} = \textit{E}\textsubscript{O} $-$ \textit{E}\textsubscript{M}; and (d) orthorhombic--tetragonal, \textit{E}\textsubscript{OT} = \textit{E}\textsubscript{T} $-$ \textit{E}\textsubscript{O}. In each panel, the left group of bars shows non-dispersion-corrected uMLFFs (light orange) alongside the DFT-PBE reference bar (red); the right group shows the dispersion-corrected models MACE-D3 and ORB-D3-v2 (light orange) alongside the DFT-PBE+D3 reference bar (red). Horizontal dashed lines indicate the DFT reference values.}
\label{fig:4}
\end{figure*}

\subsubsection{Relative ground-state energy difference}

A key quantity that can be extracted from the EOS analysis is the relative 0 K energy difference between ZrO\textsubscript{2} polymorphs, which contributes to their thermodynamic competition during phase transitions. Figure 4 summarizes the energy differences \textit{E}\textit{\textsubscript{MT}}\textit{, E}\textit{\textsubscript{TC}}\textit{, E}\textit{\textsubscript{MO}}\textit{, E}\textit{\textsubscript{OT}} obtained from the uMLFFs and compares them with DFT-PBE and DFT-PBE+D3. For the monoclinic--tetragonal E\textsubscript{MT} (Figure 4a), several models such as MatterSim and ORB lie within $\pm$2 meV/atom of the DFT-PBE reference, whereas most other uMLFFs overestimate this difference by $\sim$5 meV/atom. The dispersion-corrected MACE-D3 and ORB-D3-v2 models instead give E\textsubscript{MT} $\sim$29 and 35 meV/atom, respectively, in close agreement with our DFT-PBE+D3 reference ($\sim$26 meV/atom).

A more distinctive behavior emerges for orthorhombic-related energies. The orthorhombic phase has an energy between the M and T phases in our DFT-PBE calculations; it therefore competes directly with the tetragonal phase during the M$\rightarrow$T transition. For E\textsubscript{OT} (Figure 4d), 7net-l3i5 and MatterSim-v1.0.0-1M/5M agree well with DFT-PBE, with deviations within $\pm$2 meV/atom, whereas 7net-0, CHGNet, and MACE overestimate E\textsubscript{OT} by more than 10 meV/atom. The dispersion-corrected models MACE-D3 and ORB-D3-v2 predict the opposite ordering, with the O phase higher in energy than the T phase similar to our DFT-PBE+D3 calculations.

\subsubsection{Bulk modulus ($\kappa$)}

\begin{figure}[htbp]
\centering
\includegraphics[width=\columnwidth]{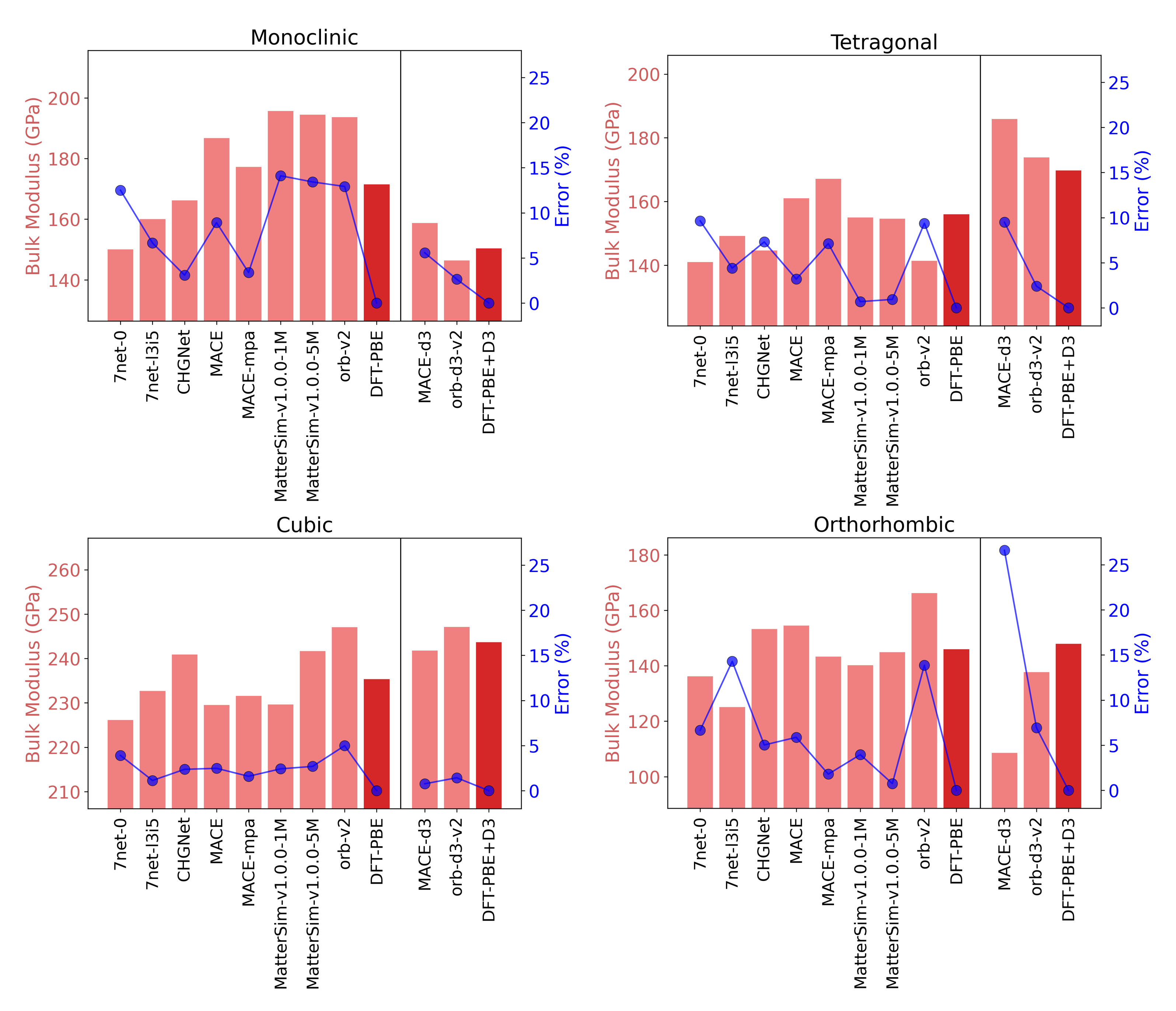}
\caption{\textbf{Bulk moduli of ZrO}\textbf{\textsubscript{2}} \textbf{polymorphs obtained from EOS fits.} Bar plots show the bulk modulus $\kappa$ for the monoclinic, tetragonal, cubic, and orthorhombic phases as predicted by the different uMLFFs. Blue symbols indicate the relative error (\%) of each model against its matched DFT reference.}
\label{fig:5}
\end{figure}

We extracted bulk moduli for the different ZrO\textsubscript{2} phases by fitting the EOS curves with the third-order Birch--Murnaghan equation of state (see Figure 5). DFT-PBE predicts the ordering $\kappa$\textsubscript{T} < $\kappa$\textsubscript{M} < $\kappa$\textsubscript{C} for the monoclinic (M), tetragonal (T), and cubic (C) phases, whereas DFT-PBE+D3 predicts $\kappa$\textsubscript{M} < $\kappa$\textsubscript{T} < $\kappa$\textsubscript{C}. All uMLFFs reproduce the ordering of their respective DFT references.

Overall, the accuracy of the bulk modulus predictions appears correlated with crystal symmetry rather than model architecture. For the low-symmetry monoclinic phase, errors can reach up to $\sim$15\%; for the tetragonal phase, most models are within $\sim$10\%; and for the high-symmetry cubic phase, errors are typically $\sim$5\%. For the orthorhombic phase, most uMLFFs achieve errors within or close to 5\%, whereas ORB-v2 and 7net-l3i5 reach 15\%.

\subsubsection{Phonon dispersion}

\begin{figure*}[tbp]
\centering
\includegraphics[width=\textwidth]{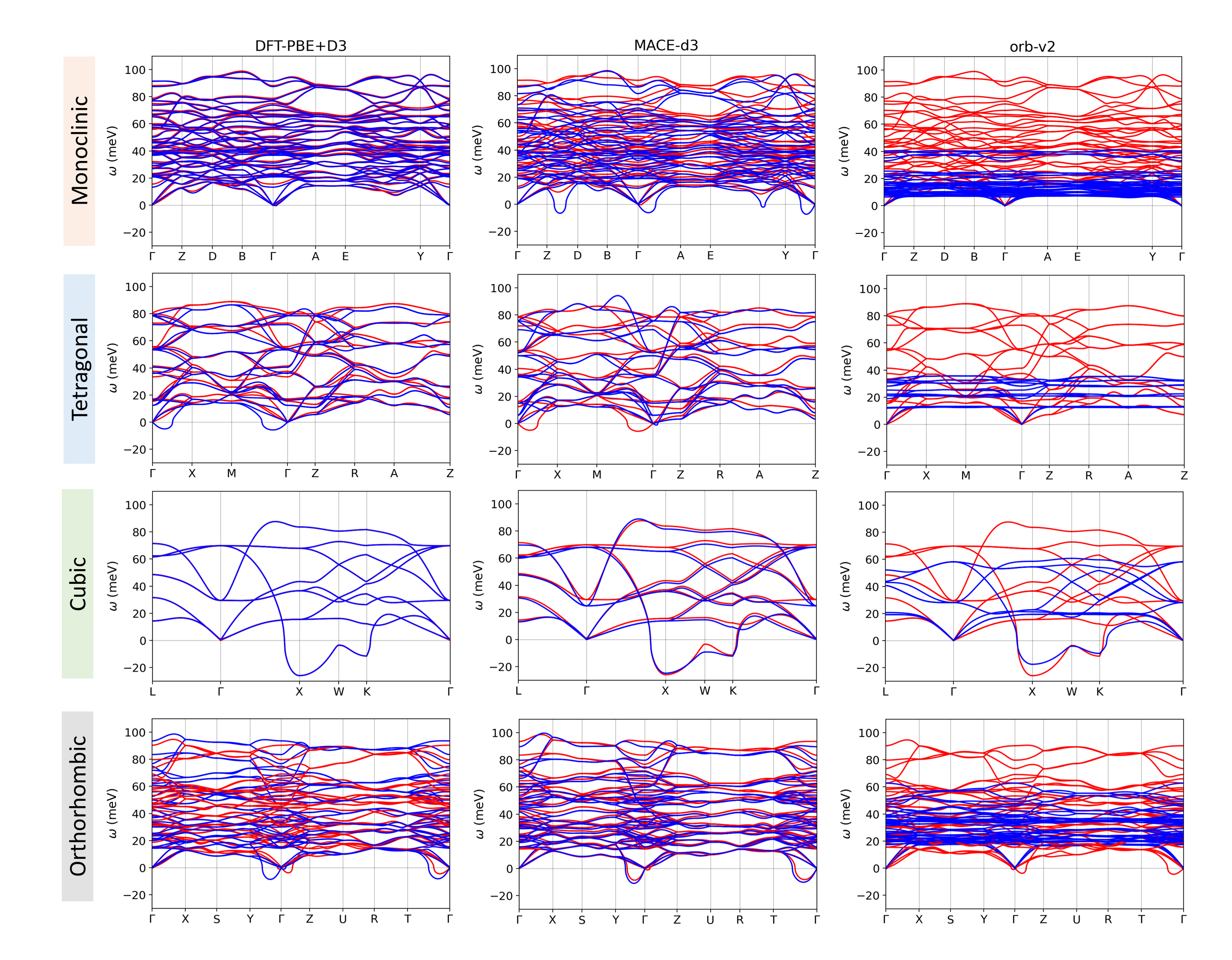}
\caption{\textbf{Phonon dispersion relations of ZrO\textsubscript{2} polymorphs compared against the matched DFT references.} Rows correspond to the monoclinic, tetragonal, cubic, and orthorhombic phases. Columns show: (left) DFT-PBE+D3 (blue) versus DFT-PBE (red) --- illustrating the intrinsic effect of the D3 correction on the DFT phonon spectrum; (middle) MACE-D3 (blue) versus DFT-PBE+D3 (red) --- the matched-functional comparison for this dispersion-corrected uMLFF; and (right) ORB-v2 (blue) versus DFT-PBE (red). Full phonon dispersions for all ten uMLFFs are provided in Supplementary Figures S3 and S4.}
\label{fig:6}
\end{figure*}

Accurately capturing the vibrational spectrum of each phase is a critical requirement for uMLFFs in studies involving finite-temperature thermodynamics and phase transitions. We computed the phonon dispersions of all ZrO\textsubscript{2} phases and compared the resulting spectra with respective DFT references. Figure 6 shows results for MACE-D3 and ORB-v2 uMLFFs as well as DFT-PBE+D3 comparison against plain DFT-PBE. Adding the D3 correction has only a small effect on the DFT phonon spectrum itself, as the maximum optical frequency $\omega$\textsubscript{max} changes by less than 5\% across all four phases at the DFT level. For the cubic phase, the DFT-PBE and DFT-PBE+D3 phonon curves are both present in Figure 6 and overlap exactly. The middle column shows that MACE-D3 reproduces the DFT-PBE+D3 phonon branches along the sampled high-symmetry paths, $\omega$\textsubscript{max} within $\pm$1.5\% for monoclinic, cubic, and orthorhombic phases, and overestimates $\omega$\textsubscript{max} by $\sim$9\% for the tetragonal phase. By contrast, ORB-v2 exhibits $\omega$\textsubscript{max} compression of $\sim$60\% for the tetragonal phase, 58\% for the monoclinic phase, and 31\% for the cubic phase. Crucially, the dispersion-corrected variant ORB-D3-v2 does not recover from this deficiency, with $\omega$\textsubscript{max} remaining 23--39\% below the DFT-PBE+D3 reference across phases (see Figure S4). All non-ORB uMLFFs reproduce $\omega$\textsubscript{max} to within $\pm$9\% of the matched DFT reference across all four phases. Li \textit{et al.} [6] also reported that ORB models yield imaginary modes near the $\Gamma$-point in PbTiO\textsubscript{3}, along with flat low-frequency branches, whereas other uMLFFs such as CHGNet, 7net-l3i5, and MACE did not show the same tendency. Consistent with these observations, Leow \textit{et al.} [12] found that ORB underperforms relative to other uMLFFs in a large-scale benchmark of $\sim$10,000 phonon calculations.

A small imaginary acoustic branch is observed near $\Gamma$ for monoclinic MACE-D3 (Figure 6) and for 7net-0 and MatterSim-v1.0.0-1M (see Figure S3) with the 2$\times$2$\times$2 supercell (same as respective DFT reference). Increasing the supercell size to 3$\times$3$\times$3 removes this feature for MACE-D3 and 7net-0, whereas a residual near-$\Gamma$ feature persists for MatterSim-v1.0.0-1M (see Figure S2). To test whether the MACE-D3 feature reflects a genuine cell-shape instability, we also performed five symmetry-free relaxations initialized with random atomic displacements of up to $\pm$0.020 \AA{} and shear perturbations of up to $\pm$0.010. All five configurations relaxed back to the monoclinic structure without triclinic drift. Therefore, the near-$\Gamma$ feature is attributable to finite-supercell truncation or interpolation of the force constants rather than to a cell-shape instability.

To quantify the agreement between uMLFF and DFT vibrational spectra, we computed the mean absolute error (MAE) in phonon frequencies following Refs.[4], [13]. Most non-ORB uMLFFs achieve mean MAEs below $\sim$5 meV for all phases, while the ORB-based models exhibit MAEs that are approximately 3--8 times higher than those of the other uMLFFs across all phases, consistent with the qualitative softening seen in Figure 6. More generally, all models display larger MAEs and wider distributions for the low-symmetry monoclinic phase and smaller errors for the high-symmetry cubic phase, mirroring the trends observed for the bulk modulus in Section 2.1.4.

\begin{figure*}[tbp]
\centering
\includegraphics[width=\textwidth]{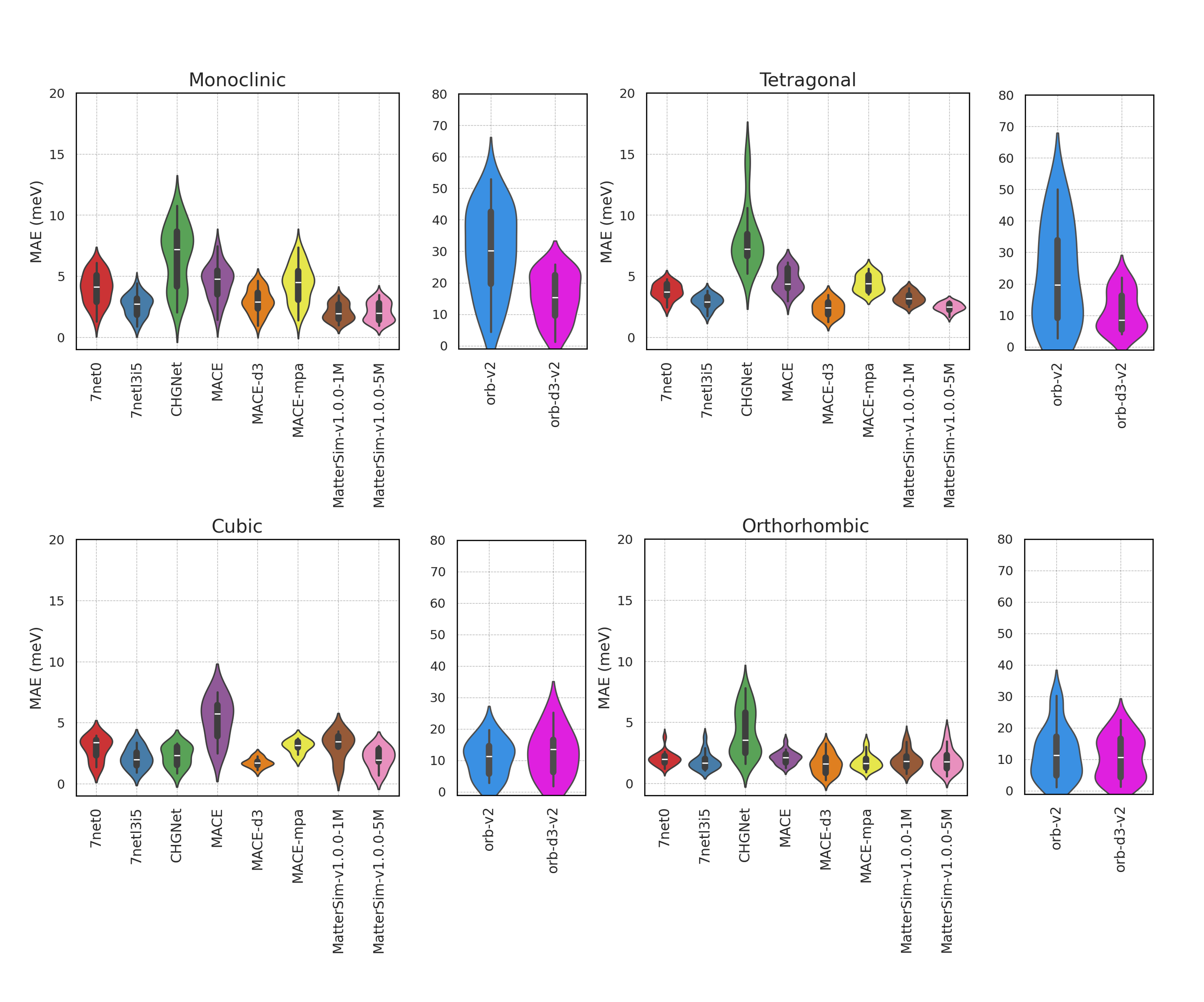}
\caption{\textbf{Mean absolute error (MAE) in phonon frequencies for ZrO}\textbf{\textsubscript{2}} \textbf{polymorphs predicted by uMLFFs.} Violin plots show the distribution of MAE values (in meV) between uMLFF and DFT phonon frequencies, evaluated over all modes and q-points, for the monoclinic, tetragonal, cubic, and orthorhombic phases. For the non-dispersion-corrected models, the reference is DFT-PBE; for the dispersion-corrected models (MACE-D3, ORB-D3-v2), the reference is DFT-PBE+D3.}
\label{fig:7}
\end{figure*}

\subsection{MD-predicted phase transitions}

Based on the MD results, we classify the uMLFFs into three tiers: (i) dynamically unstable models that cannot sustain variable-cell NPT dynamics (Section 2.2.1); (ii) dynamically stable models that predict an incorrect phase sequence involving a spurious orthorhombic intermediate (Section 2.2.2); and (iii) models that reproduce the experimental heating sequence (Section 2.2.3). The following subsections discuss each tier in turn.

\begin{figure}[htbp]
\centering
\includegraphics[width=\columnwidth]{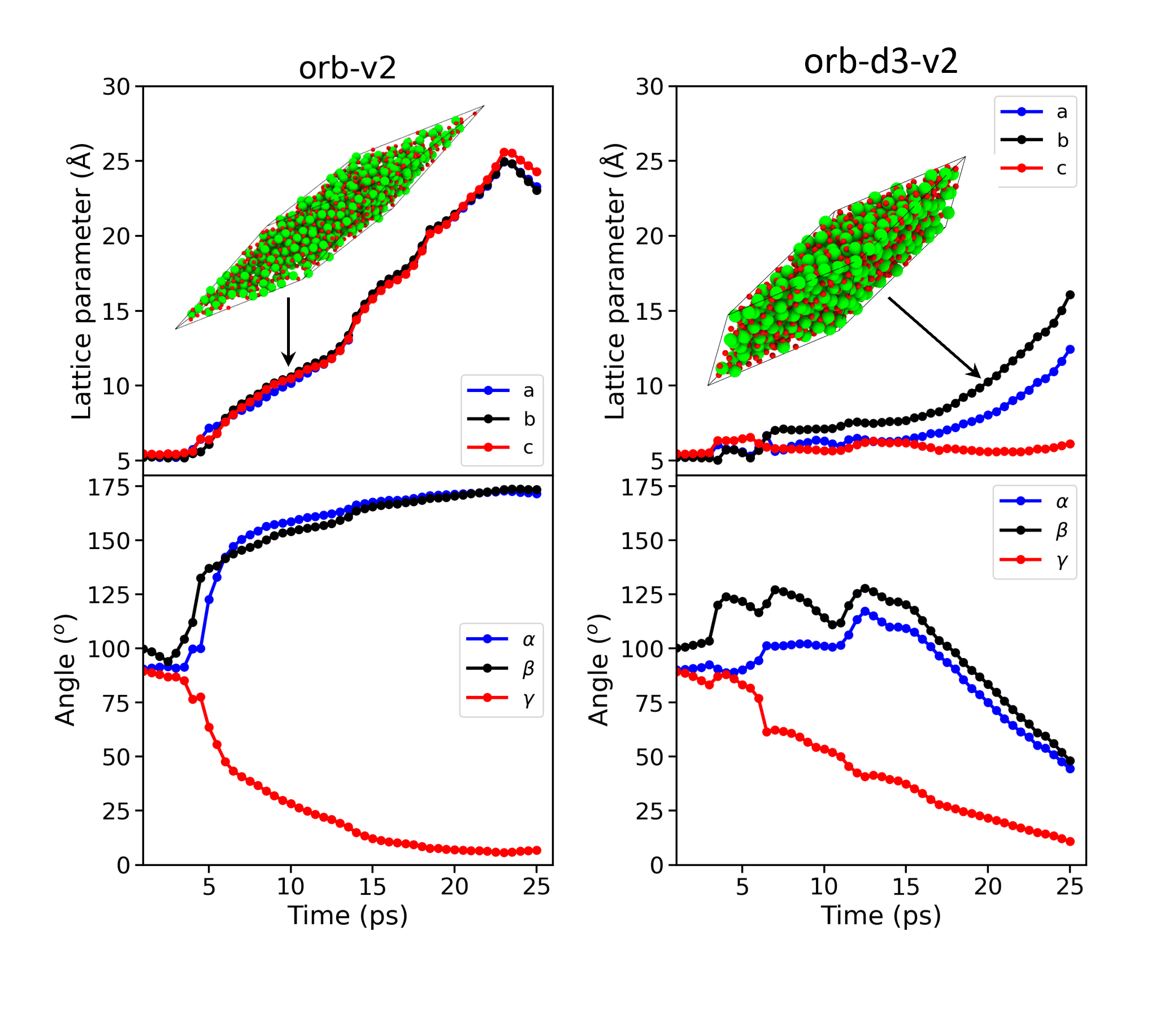}
\caption{\textbf{Time evolution of the lattice parameters (a, b, c) and cell angles ($\alpha$, $\beta$, $\gamma$) during NPT heating simulations of monoclinic ZrO}\textbf{\textsubscript{2}} \textbf{initialized at 10 K} using (left) ORB-v2 and (right) ORB-d3-v2.}
\label{fig:8}
\end{figure}

\subsubsection{Dynamically unstable models}

The ORB models (ORB-v2 and ORB-D3-v2) do not remain stable under NPT conditions (see Figure 8 and Fig. S1). Under fully flexible variable-cell conditions, rapid and unphysical lattice distortions dominated by large changes in the cell angles prevent any analysis of phase transitions in ZrO\textsubscript{2}. This behavior is consistent with the poor phonon-dispersion predictions observed for ORB in Section 2.1.5. We note, however, that Li \textit{et al.}[6] reported stable ORB-v2 dynamics for tetragonal PbTiO\textsubscript{3} and successfully simulated the tetragonal-to-cubic (T$\rightarrow$C) transition. In our case, ORB phonon dispersions for ZrO\textsubscript{2} deviate much more strongly from DFT than those reported for tetragonal PbTiO\textsubscript{3}, suggesting that ORB stability depends on the target system and degrades when the vibrational properties are not captured with sufficient fidelity.

\subsubsection{Incorrect phase sequence}

For the models that are stable under NPT conditions (7net-0, CHGNet, and MACE), the heating simulations consistently predict an incorrect sequence, M$\rightarrow$O$\rightarrow$T$\rightarrow$C$\rightarrow$L (see Figure 9, top row), in which an orthorhombic (O) phase appears between the monoclinic and tetragonal phases. In these models, the tetragonal phase appears only over a narrow temperature window of $\sim$20 K for 7net-0 and CHGNet and $\sim$80 K for MACE (Figure 9, bottom row). Consistent with this interpretation, the EOS results (Section 2.1.2, Figure 3) place the O-phase ground-state energy between those of the M and T phases for these models, making it a strong competitor during the M$\rightarrow$T transformation; correspondingly, the phase sequence observed during heating closely mirrors the ground-state energetic ordering inferred from the EOS. In cooling simulations starting from the cubic phase (Figure 9, middle row), these models show the reverse sequence O$\leftarrow$T$\leftarrow$C, and no M$\leftarrow$O transition is observed even though the O phase has imaginary modes for these uMLFFs (see phonon dispersions in Figures S3 and S4). The cooling calculations used the same system sizes as the corresponding heating calculations, and the fully flexible simulation cell allowed both lattice parameters and cell angles to evolve; therefore, the persistence of the O phase is not attributable to a smaller cooling cell or a fixed cell shape. The O phase and its implications for the phase-transition pathway are analyzed in more detail in Section 3.3.

\begin{figure*}[tbp]
\centering
\includegraphics[width=\textwidth,height=0.85\textheight,keepaspectratio]{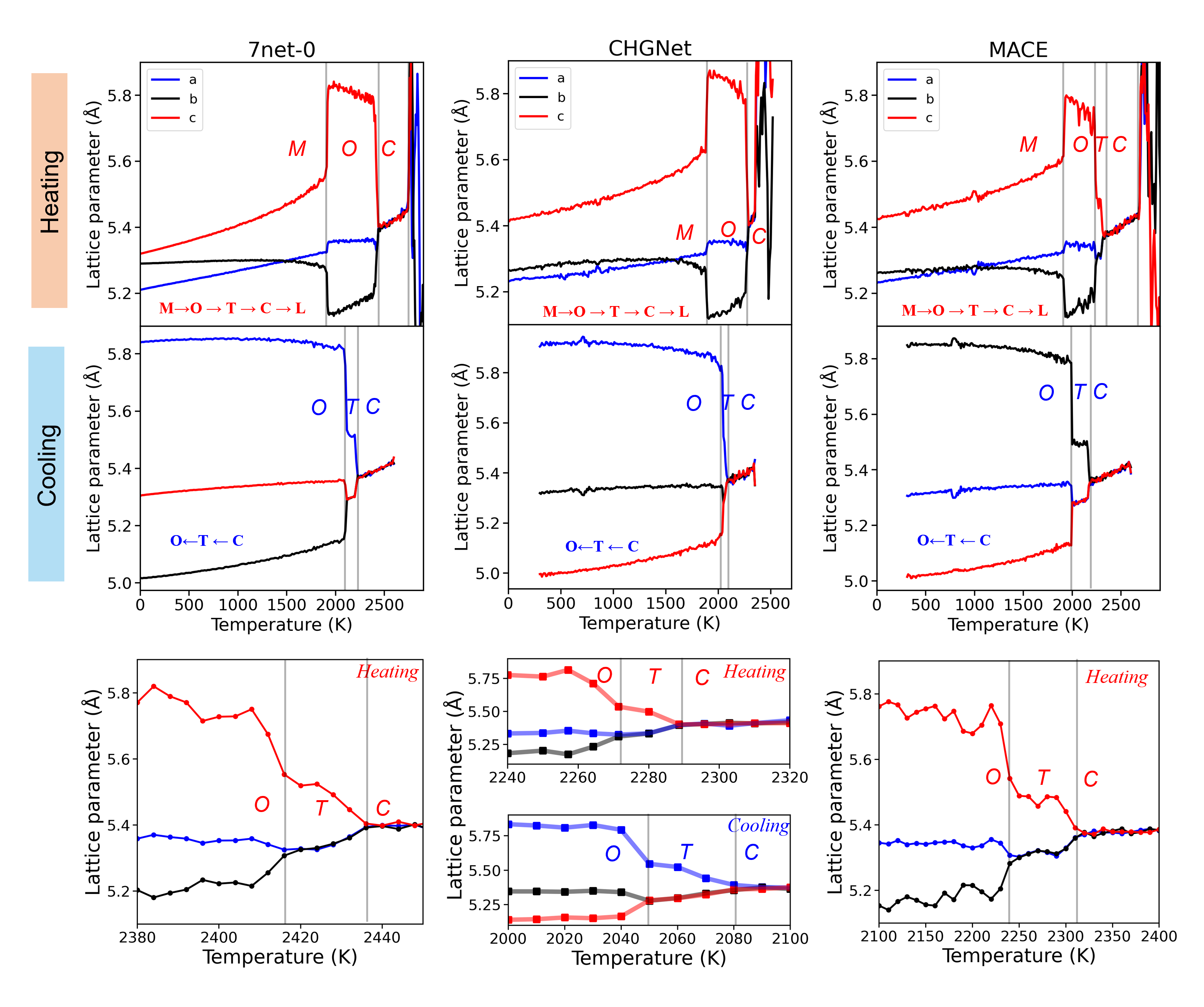}
\caption{\textbf{MD-predicted phase transitions in ZrO}\textbf{\textsubscript{2}} \textbf{for models that fail to reproduce the experimental sequence.}  \textbf{Top row:} NPT heating simulations from monoclinic ZrO\textsubscript{2} using 7net-0, CHGNet, and MACE (left to right), showing the temperature dependence of lattice parameters a, b, and c. \textbf{Middle row:} corresponding NPT cooling simulations initialized from the cubic phase obtained at the end of the heating run. Phase labels (M, O, T, C) denote temperature intervals where the instantaneous structure is classified as monoclinic, orthorhombic, tetragonal, or cubic, respectively. All three models stabilize an intermediate orthorhombic phase and predict an M$\rightarrow$O$\rightarrow$T$\rightarrow$C$\rightarrow$L pathway on heating, in contrast to the experimental M$\rightarrow$T$\rightarrow$C$\rightarrow$L sequence. \textbf{Bottom row:} enlarged views of the transition region highlighting that the T phase is stable only over a narrow temperature interval.}
\label{fig:9}
\end{figure*}

\begin{figure}[htbp]
\centering
\includegraphics[width=\columnwidth]{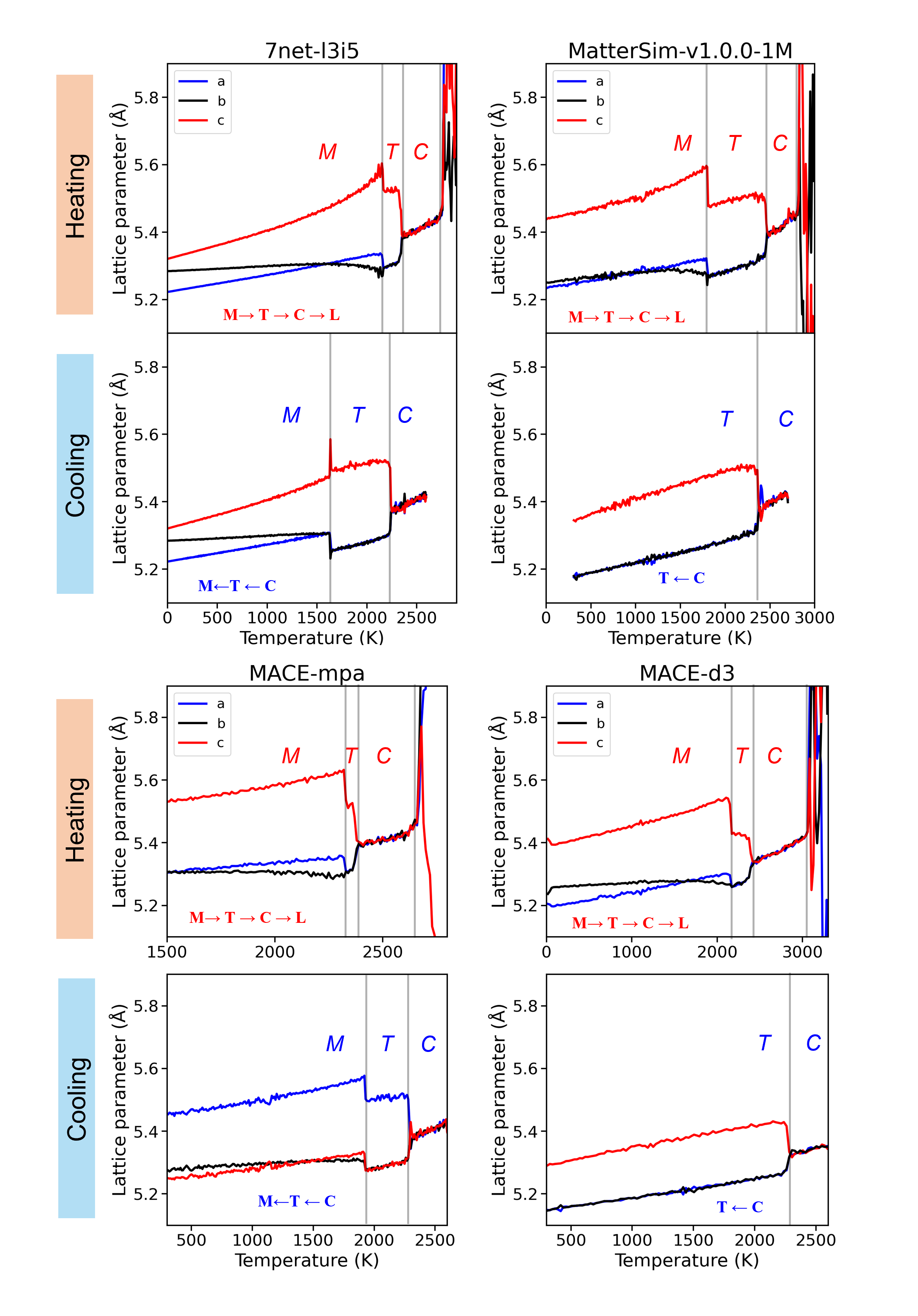}
\caption{\textbf{MD-predicted phase transitions in ZrO}\textbf{\textsubscript{2}} \textbf{for uMLFFs that reproduce the experimental sequence.} For each model, the top subpanel shows NPT heating simulations initialized from the monoclinic phase and the bottom subpanel shows NPT cooling simulations initialized from the cubic phase, with the evolution of lattice parameters a, b, c plotted versus temperature. Results are shown for (a) 7net-l3i5, (b) MatterSim-v1.0.0-1M, (c) MACE-mpa, and (d) MACE-D3.}
\label{fig:10}
\end{figure}

\subsubsection{Correct phase sequence}

We now turn to the uMLFFs (7net-l3i5, MatterSim-v1.0.0-1M, MACE-mpa, and MACE-D3) that reproduce the experimentally established correct M$\rightarrow$T$\rightarrow$C$\rightarrow$L heating sequence upon NPT heating simulations (see Figure 10). In cooling simulations starting from the cubic phase, 7net-l3i5 and MACE-mpa exhibit the full reverse sequence M$\leftarrow$T$\leftarrow$C, showing both the T$\leftarrow$C and M$\leftarrow$T transitions. In contrast, MatterSim-v1.0.0-1M and MACE-D3 show only T$\leftarrow$C and remain tetragonal at the lowest simulated temperatures, similar to the \textquotedbl{}on-the-fly\textquotedbl{} VASPML-SVD reference [7]. The persistence of the T phase during cooling is consistent with these models predicting stable tetragonal phonon modes.

These results demonstrate that some pretrained uMLFFs can capture the correct phase-transition sequence in zirconia without any system-specific retraining. This performance exceeds that of many earlier classical force fields, which generally required material-specific parameter optimization to reproduce complex structural phase behavior.

\subsubsection{Transition temperatures and phase stability}

Figure 11 summarizes (i) the transition temperatures and (ii) the temperature windows over which each polymorph is stabilized during heating and cooling for different uMLFFs. For the models that reproduce the correct heating sequence, the largest discrepancies occur for the low-temperature M$\rightarrow$T transition. The predicted M$\rightarrow$T transition temperatures exceed the experimental and VASPML-SVD values by more than 400 K, ranging from 1800 to 2200 K for the uMLFFs compared with 1400 K in experiment [14] and 1493 K for VASPML-SVD [7]. Consequently, the temperature-stability window of the T phase is substantially reduced: it spans $\sim$1100 K experimentally, but only $\sim$210 K for 7net-l3i5, 670 K for MatterSim-v1.0.0-1M, and $\sim$50 K for MACE-mpa. Agreement improves at higher temperatures: deviations are within 150 K for the T$\rightarrow$C transition and within $\sim$100 K for the C$\rightarrow$L transition.

For the models that stabilize an intermediate orthorhombic phase, the tetragonal phase is only marginally stabilized: during heating, its stability window is $\sim$20--80 K, whereas during cooling it is $\sim$30--200 K. The MD simulations also exhibit pronounced thermal hysteresis between heating and cooling: 400--600 K for the M$\leftrightarrow$T transition versus $\sim$100 K for T$\leftrightarrow$C. Such hysteresis is common in MD studies and arises from the fast heating and cooling rates (2 K/ps) and defect-free supercells, which suppress heterogeneous nucleation. If an accurate phase-transition temperature were the primary objective, a two-phase simulation at constant enthalpy could be used so that the system self-equilibrates near the transition temperature. The intended purpose here, however, is to benchmark phase-transition sequences and model stability rather than to determine high-accuracy transition temperatures.

\begin{figure}[htbp]
\centering
\includegraphics[width=\columnwidth]{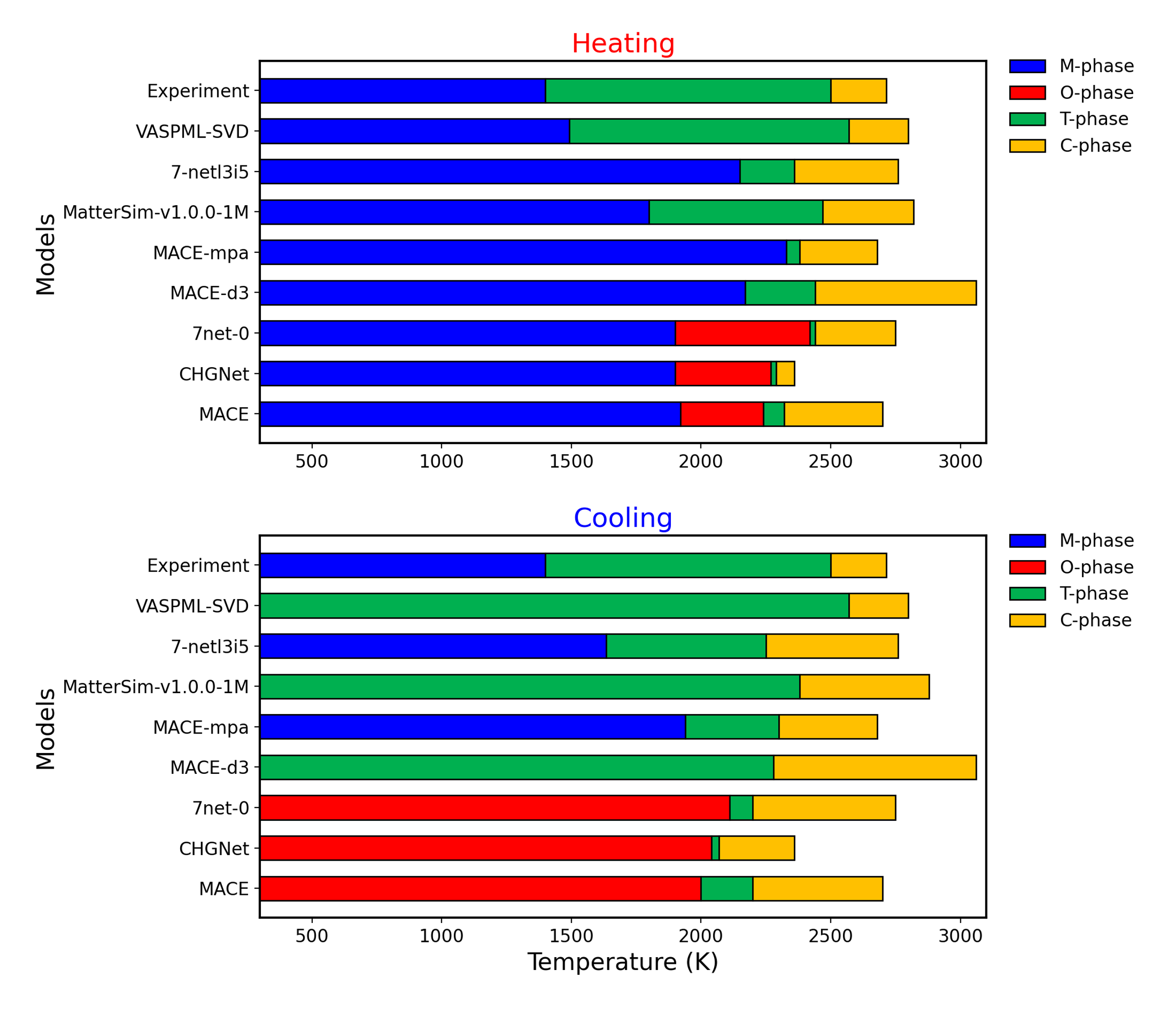}
\caption{\textbf{Comparison of MD-predicted phase-transition temperatures and phase-stability windows for ZrO}\textbf{\textsubscript{2}}\textbf{.} Phase-transition analysis for (a) heating simulations initialized from the monoclinic phase and (b) cooling simulations initialized from the cubic phase. For each uMLFF, the figure summarizes the transition temperatures and the temperature ranges over which the monoclinic (M), orthorhombic (O), tetragonal (T), and cubic (C) phases are stabilized, and compares these results with experiment and the VASPML-SVD reference. Numerical values are provided in Supplementary Section S1 (Tables S1).}
\label{fig:11}
\end{figure}

\subsubsection{Volume expansion and transition order}

Experimentally, the M$\rightarrow$T transition in ZrO\textsubscript{2} is first order, as evidenced by a clear volume discontinuity near $\sim$1400 K, whereas the T$\rightarrow$C transition is second order and is reflected by a continuous change in \textit{V}(\textit{T}) across the transition (Figure 12, experiment). Because both the experimental data and the prior VASPML-SVD report provide \textit{V}(\textit{T}) for direct comparison, we use the presence or absence of a volume discontinuity as the common diagnostic here rather than an enthalpy discontinuity. For the models that predict the incorrect phase sequence (see Figure 12a), the low-temperature volumes are systematically larger than the experimental values, consistent with the ground-state overestimation of lattice parameters discussed in Section 2.1.1. In these models, the M$\rightarrow$O transition shows a first-order-like volume signature with a volume increase, while the subsequent O$\rightarrow$T transition also shows a discontinuous volume change but with a decrease. The T$\rightarrow$C transition retains a second-order-like volume signature, appearing as a continuous volume increase. These assignments are based on \textit{V}(\textit{T}); a more rigorous thermodynamic determination of transition order would additionally examine the enthalpy for a discontinuity. These three models produce similar \textit{V}(\textit{T}) trajectories, suggesting shared systematic behavior that may reflect their common MPtraj-based training data.

For the models that reproduce the experimental heating sequence (see Figure 12b), the V(T) curves start at larger volumes than experiment at low temperature, consistent with their ground-state volumes. MACE-D3 starts at a slightly smaller equilibrium volume than the other three, reflecting the contraction induced by its D3 dispersion correction. In comparison with the experimental and VASPML-SVD V(T) references, these models reproduce a first-order-like M$\rightarrow$T volume discontinuity and a second-order-like continuous T$\rightarrow$C volume signature.

\begin{figure*}[tbp]
\centering
\includegraphics[width=\textwidth]{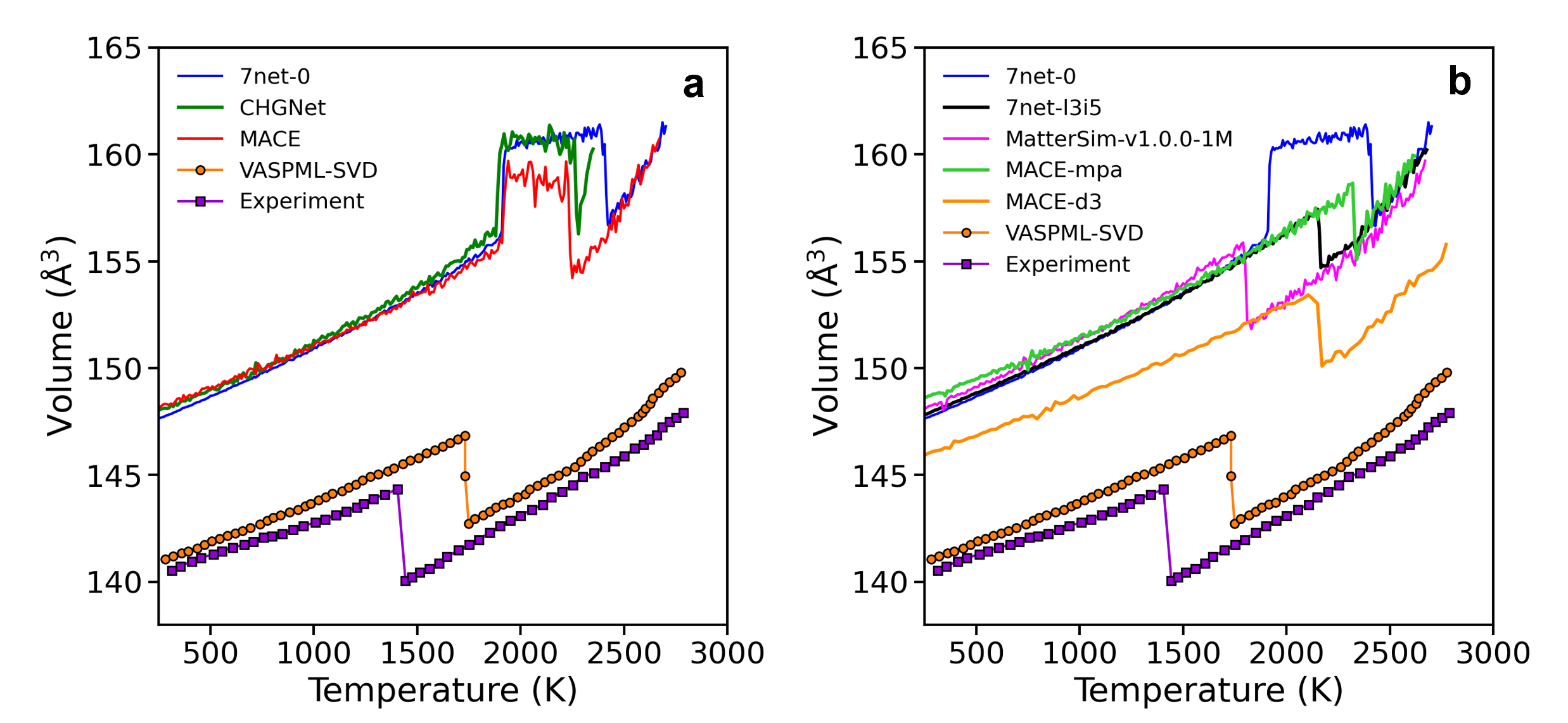}
\caption{\textbf{Volume--temperature behavior and transition order from NPT heating simulations of ZrO}\textbf{\textsubscript{2}}\textbf{.} Total simulation-cell volume normalized by the number of replicated unit cells, V(T) , during NPT heating, compared with experiment and on-the-fly VASPML-SVD. (a) Models that predict an intermediate orthorhombic phase and follow M$\rightarrow$O$\rightarrow$T$\rightarrow$C$\rightarrow$L (7net-0, CHGNet, MACE). (b) Models that reproduce the experimental sequence M$\rightarrow$T$\rightarrow$C$\rightarrow$L (7net-l3i5, MatterSim-v1.0.0-1M, MACE-mpa, MACE-D3); 7net-0 is shown as a reference. The experimental M$\rightarrow$T transition is first order (volume discontinuity), whereas T$\rightarrow$C is second order (continuous V(T)).}
\label{fig:12}
\end{figure*}

\section{DISCUSSION}

In this section, we analyze the correct and incorrect phase-transition sequences obtained in Sections 2.2.1--2.2.5 and relate them to the ground-state benchmarks in Section 2.1. Our goal is to identify which ground-state metrics are most predictive of finite-temperature behavior and to assess whether failures in MD arise from limitations in the training data and/or model design.

\subsection{Stability of uMLFF models}

Here, we define model stability as the ability to sustain NPT dynamics beginning from the monoclinic (M) phase at 10 K without unphysical lattice collapse or runaway cell deformations. Under this criterion, all tested models except the ORB potentials remain stable (Figures 8--10). The models considered here differ in whether forces are predicted directly by a force head or obtained as analytic gradients of a learned energy. ORB uses direct force prediction, but this distinction alone does not establish the cause of its instability, because other architectures may also employ dedicated force outputs. We therefore do not attribute the observed behavior solely to the force-prediction strategy. Instead, the direct evidence in this study is the large error and systematic softening in the ORB phonon spectra. Combined with the phonon analysis in Section 2.1.5, this result shows that phase-transition studies require not only accurate relative polymorph energetics but also a sufficiently faithful vibrational response to support stable variable-cell dynamics.

\subsection{Influence of training data and model architecture}

Sections 2.2.1--2.2.2 show that models trained on broadly similar datasets can nonetheless yield qualitatively different phase-transition pathways in ZrO\textsubscript{2}. Here we compare pairs of closely related models to clarify the relative roles of training data, architectural capacity, and training choices.

For SevenNet, the 7net-0 and 7net-l3i5 pretrained models are trained on the same dataset (MPtraj) but differ in architectural resolution: 7net-l3i5 increases the maximum spherical-harmonic degree from \textit{l}\textsubscript{max} = 2 to \textit{l}\textsubscript{max} = 3 and uses five radial basis functions (``i5'') in the local-neighborhood expansion. With the training data held fixed, this architectural change alters the predicted heating sequence from the incorrect M$\rightarrow$O$\rightarrow$T$\rightarrow$C$\rightarrow$L pathway (7net-0) to the experimentally consistent M$\rightarrow$T$\rightarrow$C$\rightarrow$L pathway (7net-l3i5). This improvement, however, comes at an increased computational cost of approximately fourfold when the maximum spherical-harmonic degree is increased.

A similar trend is observed within the MACE family. MACE and MACE-mpa share the same underlying architecture, while MACE-mpa incorporates updated training choices (updated hyperparameters and trained weights). This update likewise changes the predicted phase-transition sequence from M$\rightarrow$O$\rightarrow$T$\rightarrow$C$\rightarrow$L (MACE) to M$\rightarrow$T$\rightarrow$C$\rightarrow$L (MACE-mpa).

Mechanistically, these changes in finite-temperature behavior are consistent with improvements in the ground-state description: the model variants that recover the correct phase sequence also bring key relative energy differences $-$ most notably the orthorhombic--tetragonal difference E\textsubscript{OT} closer to the DFT-PBE reference (Figure 4), reducing the tendency to stabilize the competing orthorhombic polymorph. Together, these comparisons indicate that even with similar training data, higher representational capacity in SevenNet and improved training and hyperparameter choices in MACE can change the learned relative energetics among competing polymorphs enough to determine whether the correct phase-transition pathway is recovered.

\subsection{Metastable phase analysis}

As noted in Section 2.2.2, several models (7net-0, CHGNet, and MACE) stabilize an additional orthorhombic (O) polymorph between the monoclinic (M) and tetragonal (T) phases during MD. Here we examine the structural identity of that O phase and why it becomes competitive along the M$\rightarrow$T pathway.

Although this O phase is not present in the Materials Project database used by several of these pretrained models, we identified a corresponding orthorhombic ZrO\textsubscript{2} structure in the NIST JARVIS database (JVASP-151841). After applying a rotational-and-translational symmetry operation, the orthorhombic structure produced by 7net-0, CHGNet, and MACE matches the JARVIS entry JVASP-151841 in space group and atomic connectivity (Figure S6), while the powder XRD comparison is shown in Figure S7.

Structurally, the O phase can be generated from the tetragonal polymorph by shifting the body-centered Zr atoms away from their symmetry positions, accompanied by distortions of the oxygen sublattice (Figure 13b). This suggests that the low-symmetry O phase may represent a plausible metastable intermediate along the M$\rightarrow$T transformation pathway and can be energetically competitive with the T phase. The associated volume evolution and energy-per-atom changes across the orthorhombic--tetragonal transition are shown in Figure S5. We quantify how errors in the orthorhombic--tetragonal energy differences E\textsubscript{OT} correlate with stabilization of this O phase in the next Section 3.4.

\begin{figure}[htbp]
\centering
\includegraphics[width=\columnwidth]{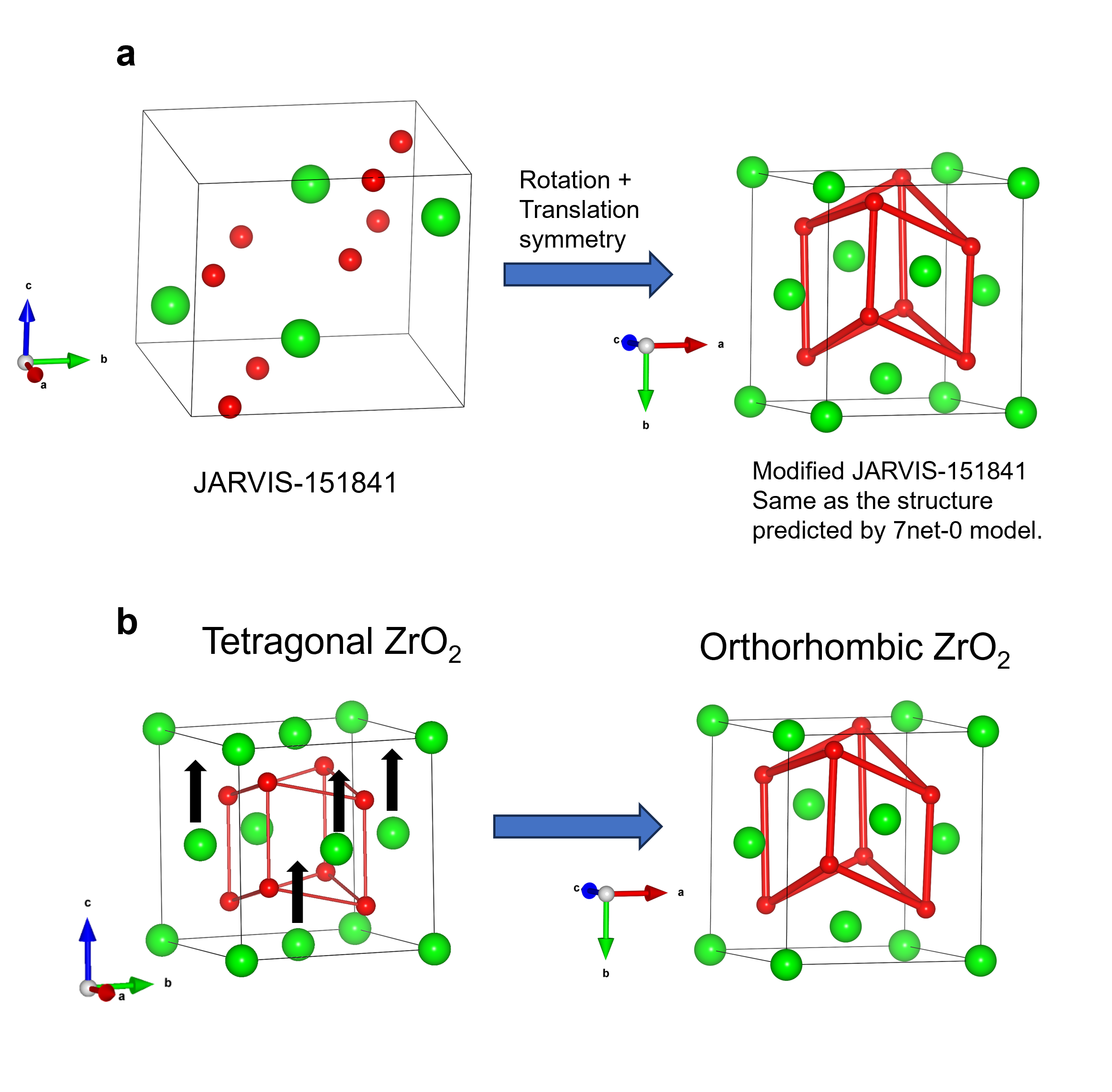}
\caption{\textbf{Identification and structural origin of the} \textbf{\textit{O}}\textbf{-ZrO}\textbf{\textsubscript{2}} \textbf{phase observed in uMLFF-driven MD.} (a) Rotational and translational symmetry operations applied to the JARVIS \textit{O}-ZrO\textsubscript{2} structure (JVASP-151841) to obtain a symmetry-equivalent unit cell for direct comparison with the O phase predicted in MD using 7net-0, CHGNet, and MACE. (b) Schematic illustration of the transformation from tetragonal to orthorhombic ZrO\textsubscript{2}, highlighting the displacement of the body-centered Zr atoms from their symmetry positions and the accompanying distortion of the oxygen sublattice. Green and red spheres denote Zr and O atoms, respectively.}
\label{fig:13}
\end{figure}

\subsection{Ground-state energetics rationalize MD outcomes}

Sections 2.1.2 and 2.2 establish the two ingredients needed to rationalize the MD phase behavior: the 0 K energetics from the EOS analysis and the temperature-dependent transition sequences from NPT MD. To quantify model-to-DFT differences in this competition at 0 K, we define the deviation $\Delta$ from the DFT-PBE reference as evaluated at 0 K. This analysis is restricted to models that maintain the DFT-PBE-like ordering E\textsubscript{O}<E\textsubscript{T} (i.e., E\textsubscript{OT}>0).

Models that sustain stable NPT dynamics and have $\Delta$ $\leq$ $\sim$6 meV/atom reproduce the correct heating sequence. In contrast, models with $\Delta$ $\geq$ $\sim$11 meV/atom stabilize an intermediate O phase. The transition between these two regimes therefore lies in the $\sim$6--10 meV/atom range. Within this 0 K picture, overestimating the O--T energy difference makes the O phase too competitive along the M$\rightarrow$T transformation pathway. As the system leaves the monoclinic basin, it can pass through an O-like configuration before reaching the T phase. Models with $\Delta$ $\leq$ $\sim$6 meV/atom also achieve low phonon MAEs (<5 meV on average) without systematic softening or spurious compression of the vibrational spectrum (Figures 6 and 7). By contrast, models with $\Delta$ $\geq$ $\sim$11 meV/atom exhibit much larger phonon MAEs and pronounced low-frequency compression. These systematic errors in relative phase energetics and vibrational properties correlate directly with the appearance of an artificial M$\rightarrow$O$\rightarrow$T pathway or outright MD instability. This result underscores that accurate ground-state energy differences and phonon dispersions are prerequisites for reliable phase-transition predictions by uMLFFs.

The 0 K analysis above neglects finite-temperature contributions. As a limited check of whether thermal sampling changes the energetic trend, we compared trajectory-averaged O--T potential-energy splits within the model-specific transition intervals identified in the MD runs with their 0 K EOS values (Supplementary Section S5). This shift is temperature dependent, and the trajectory-averaged potential-energy difference is not a free-energy difference because it does not include the entropic contribution. For the models that follow the experimental sequence, the sampled finite-temperature values reduce E\textsubscript{OT} by only 2--4 meV/atom relative to the 0 K values, which is smaller than the 6--11 meV/atom band separating the correct- and incorrect-sequence models; for 7net-0, the deviation from DFT-PBE persists in the sampled transition interval. This comparison is therefore used only to confirm that thermal changes over the sampled intervals do not overturn the 0 K energetic trend, rather than as a substitute for a finite-temperature free-energy calculation. The 0 K $\Delta$ consequently remains a useful pre-flight predictor of the observed finite-temperature phase-transition behavior.

For MACE-D3 and ORB-D3-v2, inheriting the DFT-PBE+D3 ordering E\textsubscript{M}<E\textsubscript{T}<E\textsubscript{O}<E\textsubscript{C} yields E\textsubscript{OT}<0, placing the O phase energetically above T rather than between M and T. Under this ordering, the O configuration is no longer a steppingstone along the M$\rightarrow$T transformation on heating. Thermal activation out of the M basin finds T directly because T is energetically lower than O. Thus, the inverted ordering removes the orthorhombic trap that affects 7net-0, CHGNet, and MACE.

\section{CONCLUSION}

We benchmarked ten pretrained universal machine-learning force fields (uMLFFs) for ZrO\textsubscript{2} by combining 0 K ground-state calculations with variable-cell NPT MD heating and cooling simulations. Four models (7net-l3i5, MatterSim-v1.0.0-1M, MACE-mpa, MACE-D3) reproduced the experimental M$\rightarrow$T$\rightarrow$C$\rightarrow$L heating sequence; three (7net-0, CHGNet, MACE) stabilized an orthorhombic intermediate and predicted M$\rightarrow$O$\rightarrow$T$\rightarrow$C$\rightarrow$L; and two (ORB-v2, ORB-D3-v2) failed to sustain stable NPT dynamics altogether.

These outcomes support practical screening metrics for choosing models for finite-temperature phase-transition studies with pretrained uMLFFs: relative polymorph energetics should agree with the matched DFT reference to within about 6 meV/atom, and the phonon dispersion should not show systematic softening or low-frequency compression. In our benchmarks, orthorhombic--tetragonal energy errors above $\sim$10 meV/atom consistently produced an O-phase intermediate, whereas the ORB models failed despite reasonable energy ordering because their phonon spectra were strongly compressed. Together, these two diagnostics correctly classified all ten models at far lower cost than a full NPT campaign.

The MACE-D3 result warrants a specific caveat for practitioners applying this protocol. MACE-D3 reproduces the M$\rightarrow$T$\rightarrow$C$\rightarrow$L sequence not by matching DFT-PBE polymorph energetics, but because its DFT-PBE+D3 training reference places the orthorhombic phase above the tetragonal phase. Accordingly, the MACE-D3 result should not be read as a general endorsement of dispersion-corrected uMLFFs for phase-transition studies. ORB-D3-v2, trained against the same class of reference, still fails dynamically (Section 2.2.1). When applying the two-diagnostic protocol above, the DFT reference must be chosen consistently with the uMLFF's training functional. Otherwise, apparent energy errors may reflect a functional or correction mismatch rather than a model deficiency.

The comparison nevertheless shows that D3 can materially alter the relative stability of competing ZrO\textsubscript{2} polymorphs: at the matched DFT-PBE+D3 level, the O--T ordering is reversed relative to DFT-PBE, and MACE-D3 inherits this ordering. This sensitivity indicates that interactions represented by the dispersion correction affect the delicate energetic balance between the tetragonal and orthorhombic structures, although the present calculations do not establish that van der Waals interactions are universally required for ZrO\textsubscript{2}. A related architectural result is that increasing SevenNet's angular resolution from \textit{l}\textsubscript{max} = 2 to \textit{l}\textsubscript{max} = 3 improves the relative polymorph energetics and recovers the experimental phase sequence. This is consistent with higher-order spherical harmonics better resolving the orientation-dependent local environments associated with the lower-symmetry phases, but it should be interpreted as an empirical result of this benchmark rather than as an a priori requirement for all zirconia models.

Even when the exact phase sequence is not recovered, uMLFF predictions carry useful screening value: in the present case, all dynamically stable models correctly identified ZrO\textsubscript{2} as a multi-transition material, information that is experimentally time-consuming to establish. As training datasets expand and architectures improve, uMLFF accuracy for finite-temperature properties should continue to improve. To accelerate this progress, we recommend that the community establish standardized phase-transition benchmarks --- analogous to MatBench Discovery [3] for static properties --- that systematically evaluate uMLFF fidelity for dynamic-property prediction across diverse material classes. The two diagnostics identified here, relative polymorph energetics and phonon-dispersion fidelity against the matched DFT reference, provide a practical starting point for such benchmarks and an immediate pre-flight screening tool for practitioners applying pretrained uMLFFs to phase-transition prediction in complex oxides.

\section{METHODS}

\subsection{Universal machine-learning force-field models (uMLFFs)}

We evaluated ten pretrained universal machine-learning force fields (uMLFFs) to compute ZrO\textsubscript{2} ground-state properties and to perform molecular-dynamics (MD) simulations. All models were accessed through the CHIPS-FF infrastructure [4] using the corresponding implementations and released checkpoints.

For clarity, we use the following shorthand for training datasets: MPtrj denotes the Materials Project GGA-PBE relaxation trajectories; Alex denotes the Alexandria GGA-PBE dataset [16]; and sAlex is a subsampled version of Alexandria. The models evaluated are: CHGNet [2] (v0.3.8, trained on MPtrj); SevenNet [9] (7net-0: \textit{l}\textit{\textsubscript{max}}=2, 11 Jul 2024; and 7net-l3i5: \textit{l}\textit{\textsubscript{max}}=3, 12 Dec 2024; both trained on MPtrj, with 7net-l3i5 approximately 4$\times$ slower than 7net-0); ORB [11] (v0.4.1: orb-v2 trained on MPtrj+Alex, and orb-d3-v2 trained on MPtrj+Alex with an additional D3 dispersion-correction reference set [17], [18]); MACE [10] (v0.3.10: MACE-MP-0 medium, referred to here as MACE; MACE-MPA-0 medium, referred to as MACE-mpa; and MACE-MP-0 medium with an additional D3 dispersion-correction reference set [17], [18], referred to as MACE-d3; the MACE variants are trained on MPtrj, with MACE-mpa additionally trained on sAlex); and MatterSim-v1[8] (MatterSim-v1.0.0-1M and MatterSim-v1.0.0-5M), trained on a curated dataset of $\sim$17 million structures generated via first-principles active learning, spanning temperatures of 0--3000 K and pressures of 0--1000 GPa. All uMLFFs considered here predict energies, forces, and stresses; CHGNet additionally predicts magnetic moments.

\subsection{Simulation details}

\subsubsection{Ground-state properties}

For all uMLFFs and for DFT, we performed geometry-relaxation calculations of ZrO\textsubscript{2} polymorphs starting from initial unit-cell structures obtained from the Materials Project [19]. All uMLFF calculations were carried out using the Atomic Simulation Environment (ASE v3.27.0)[20], and DFT calculations were performed using the Vienna Ab initio Simulation Package (VASP v5.4.4)[21].

To preserve the target space-group symmetry of the unit cell during relaxation, we employed the ASE symmetrizer FrechetCellFilter while allowing volume changes. Structural optimizations were performed using the FIRE algorithm as implemented in ASE, with a force-convergence criterion of 10\textsuperscript{-5} eV/\AA{} for all uMLFFs. For DFT relaxations, we used the GGA-PBE exchange-correlation functional with a plane-wave cutoff energy of 560 eV, electronic self-consistency convergence of 10\textsuperscript{-8} eV, and ionic convergence of 10\textsuperscript{-5} eV/\AA{}, with crystal symmetry maintained while all cell vectors and angles were free to relax. For the DFT-PBE+D3 reference and the D3-corrected uMLFF models (MACE-D3, ORB-D3-v2), Grimme's D3 dispersion correction with Becke--Johnson damping (D3(BJ)) [17], [18] was incorporated via the torch-dftd3 library [22]. To ensure methodological consistency between DFT and uMLFF D3-corrected calculations, the dispersion correction was applied externally at each ionic step: the base interatomic forces (from VASP GGA-PBE or the respective uMLFF) and the D3 forces were evaluated independently and summed as \textit{F}\textsubscript{total} = \textit{F}\textsubscript{base} + \textit{F}\textsubscript{D3}; the combined force was then passed to the ASE FIRE optimizer to update atomic positions and cell degrees of freedom. Lattice parameters and ground-state energies were extracted from the relaxed structures.

Equations of state (EOS) were computed by scaling the equilibrium volume by $\pm$3\%. At each scaled volume, the cell shape and ionic positions were relaxed at fixed volume using FrechetCellFilter with constant volume in ASE (equivalent to VASP ISIF=4), while cell vectors and angles were free to adjust. The same external D3 force-summation protocol was applied for DFT-PBE+D3 and D3-corrected uMLFF EOS calculations. The resulting energy--volume data were fit to the Birch--Murnaghan equation of state, from which the 0 K bulk modulus was obtained.

Phonon dispersions for DFT-PBE, DFT-PBE+D3, and the uMLFFs were computed using the PHONOPY package [23] along high-symmetry Brillouin-zone paths for each crystal structure. A 2$\times$2$\times$2 supercell was used for all ZrO\textsubscript{2} polymorphs, and force constants were obtained via the finite-displacement (FD) method with a displacement amplitude of 0.01 \AA{}. $\Gamma$-centered Monkhorst--Pack k-meshes were used for both the unit-cell and supercell DFT calculations, and the external D3 force-summation protocol was applied to each displaced-supercell single-point calculation for DFT-PBE+D3. Non-analytical corrections (NAC) for long-range polar interactions were not applied, so that DFT and uMLFF dispersions are compared on equal footing (the uMLFFs do not provide Born effective charges); LO-TO splitting at $\Gamma$ is therefore absent from all dispersions shown. Agreement between uMLFF and DFT phonons was quantified using the mean absolute error, $\mathrm{MAE} = \frac{1}{N_q N_m}\sum_{q,m}\left|\hbar\omega_{q,m}^{\mathrm{uMLFF}} - \hbar\omega_{q,m}^{\mathrm{DFT}}\right|$, where $N_q$ is the number of sampled wavevectors along the high-symmetry \textbf{q}-path and $N_m$ is the number of phonon modes. The reference frequencies $\omega_{q,m}^{\mathrm{DFT}}$ were taken from DFT-PBE for standard uMLFFs and from DFT-PBE+D3 for D3-corrected models (MACE-D3 and ORB-D3-v2), following Refs. [4], [13].

\subsubsection{Molecular dynamics simulation}

We performed isothermal-isobaric (NPT) ensemble heating and cooling simulations using the Atomic Simulation Environment (ASE) and the Large-scale Atomic/Molecular Massively Parallel Simulator (LAMMPS) [24]. All models except 7net-l3i5 were simulated using ASE; 7net-l3i5 was simulated using LAMMPS. In ASE, NPT dynamics employed the isotropic Martyna--Tobias--Klein (MTK) algorithm, and in LAMMPS we used a Parrinello--Rahman-type barostat. In both implementations, the simulation cell was fully flexible, allowing both cell angles and lattice parameters to evolve.

All MD simulations were carried out under periodic boundary conditions (PBCs) at a target pressure of 0 GPa with a 1 fs integration time step. Thermostat and barostat damping times were set to 100 and 1000 time steps, respectively. Heating was performed at an effective rate of 2 K$\cdot$ps\textsuperscript{-1}. In LAMMPS, the temperature was ramped continuously, whereas in ASE it was increased in 10 K increments and equilibrated for 5 ps, corresponding to the same effective heating rate of 2 K$\cdot$ps\textsuperscript{-1}. Heating simulations were initialized from the monoclinic ZrO\textsubscript{2} phase using a 5$\times$5$\times$5 supercell (1500 atoms) and performed over 10--3000 K to probe phase transitions. Cooling simulations were initialized from the cubic phase obtained at the end of the heating run and cooled at 2 K/ps down to room temperature. During heating and cooling, the 0.5 ps time-averaged cell parameters a, b, c, $\alpha$, $\beta$, and $\gamma$ were monitored and used for phase classification.

\section*{DECLARATION}

\textbf{Conflict of interest.} The authors declare that they have no conflict of interest.

\section*{DATA AVAILABILITY}

All data supporting the findings of this study are included within the manuscript.

\section*{AUTHOR CONTRIBUTIONS}

\textbf{Khadka, Rajan}: Conceptualization (Equal); Formal analysis (Equal); Investigation (Equal); Methodology (Equal); Writing - Original Draft (Lead).

\textbf{van de Walle, Axel}: Conceptualization (supporting); Formal analysis (Equal); Investigation (supporting); Methodology (supporting); Writing - Review \& Editing (Equal).

\textbf{Tepesch, Patrick}: Conceptualization (Equal); Formal analysis (Equal); Investigation (Equal); Methodology (Equal); Writing - Review \& Editing (Equal).

\clearpage
\onecolumngrid

\raggedbottom
\renewcommand{\thefigure}{S\arabic{figure}}
\setcounter{figure}{0}
\renewcommand{\thetable}{S\arabic{table}}
\setcounter{table}{0}

\setlength{\intextsep}{0pt}
\setlength{\abovecaptionskip}{4pt}

\section*{SUPPLEMENTARY MATERIAL}

\begin{center}
{\large\bfseries Application of Pretrained Universal Machine Learning Force-Fields for Phase Transitions Prediction in ZrO\textsubscript{2}}\\[1.2ex]
Rajan Khadka\textsuperscript{1*}, Axel van de Walle\textsuperscript{2}, Patrick Tepesch\textsuperscript{1*}\\
\textit{\textsuperscript{1}}\textit{Corning Research and Development Corporation, Corning, NY 14870, USA}\\
\textit{\textsuperscript{2}}\textit{School of Engineering, Brown University, Box D, Providence, RI, 02912, USA}\\
*Corresponding authors: \href{mailto:khadkar2@Corning.com}{khadkar2@Corning.com}, \href{mailto:TepeschPD@Corning.com}{TepeschPD@Corning.com}
\end{center}

\Needspace{0.45\textheight}
\section*{S1 Transition temperatures}\nopagebreak

\begin{table}[H]
\caption{Phase-transition temperatures (K) and phase-stability windows (K) for ZrO2 from NPT MD simulations, compared with experiment and VASPML-SVD. ``Dir'' indicates heating or cooling. The $\Delta$T columns give the temperature range over which each phase is stabilized. Tier classification follows Section 2.2: Tier 1 = dynamically unstable, Tier 2 = incorrect sequence, Tier 3 = correct sequence. ORB models are listed as ``None'' because stable NPT simulations could not be obtained.}
\label{tab:S1}
\centering
\footnotesize
\setlength{\tabcolsep}{3pt}
\begin{tabular}{llccccccccc}
\toprule
 & Dir & Sequence & M$\rightarrow$T/O & O$\rightarrow$T & T$\rightarrow$C & C$\rightarrow$L & $\Delta$T\textsubscript{M} & $\Delta$T\textsubscript{O} & $\Delta$T\textsubscript{T} & $\Delta$T\textsubscript{C} \\
\midrule
Experiment & Heat & M$\rightarrow$T$\rightarrow$C$\rightarrow$L & 1400 & $-$ & 2500 & 2715 & 1400 & 0 & 1100 & 215 \\
\multirow{2}{*}{VASPML-SVD} & Heat & M$\rightarrow$T$\rightarrow$C$\rightarrow$L & 1493 & $-$ & 2570 & 2800 & 1493 & 0 & 1077 & 230 \\
 & Cool & T$\leftarrow$C & $-$ & $-$ & 2570 & 2800 & 0 & 0 & 2570 & 230 \\
 &  &  &  &  &  &  &  &  &  &  \\
\multicolumn{3}{c}{CORRECT SEQUENCE (Tier 3)} &  &  &  &  &  &  &  &  \\
\multirow{2}{*}{7net-l3i5} & Heat & M$\rightarrow$T$\rightarrow$C$\rightarrow$L & 2150 & $-$ & 2360 & 2760 & 2150 & 0 & 210 & 400 \\
 & Cool & M$\leftarrow$T$\leftarrow$C & 1635 & $-$ & 2250 & 2760 & 1635 & 0 & 615 & 510 \\
\multirow{2}{*}{MatterSim-v1.0.0-1M} & Heat & M$\rightarrow$T$\rightarrow$C$\rightarrow$L & 1800 & $-$ & 2470 & 2820 & 1800 & 0 & 670 & 350 \\
 & Cool & T$\leftarrow$C & $-$ & $-$ & 2380 & 2880 & 0 & 0 & 2380 & 500 \\
\multirow{2}{*}{MACE-mpa} & Heat & M$\rightarrow$T$\rightarrow$C$\rightarrow$L & 2330 & $-$ & 2380 & 2690 & 2330 & 0 & 50 & 300 \\
 & Cool & M$\leftarrow$T$\leftarrow$C & 1940 & $-$ & 2300 & 2690 & 1940 & 0 & 360 & 380 \\
 &  &  &  &  &  &  &  &  &  &  \\
\multicolumn{4}{c}{INCORRECT SEQUENCE (Tier 2)} &  &  &  &  &  &  &  \\
\multirow{2}{*}{7net-0} & Heat & M$\rightarrow$O$\rightarrow$T$\rightarrow$C$\rightarrow$L & 1900 & 2420 & 2440 & 2750 & 1900 & 520 & 20 & 310 \\
 & Cool & O$\leftarrow$T$\leftarrow$C & $-$ & 2100 & 2200 & 2750 & 0 & 2110 & 90 & 550 \\
\multirow{2}{*}{CHGNet} & Heat & M$\rightarrow$O$\rightarrow$T$\rightarrow$C$\rightarrow$L & 1900 & 2270 & 2290 & 2360 & 1900 & 370 & 20 & 70 \\
 & Cool & O$\leftarrow$T$\leftarrow$C & $-$ & 2040 & 2070 & 2360 & 0 & 2040 & 30 & 290 \\
\multirow{2}{*}{MACE} & Heat & M$\rightarrow$O$\rightarrow$T$\rightarrow$C$\rightarrow$L & 1920 & 2240 & 2320 & 2700 & 1920 & 320 & 80 & 380 \\
 & Cool & O$\leftarrow$T$\leftarrow$C & $-$ & 2000 & 2200 & 2700 & 0 & 2000 & 200 & 500 \\
 &  &  &  &  &  &  &  &  &  &  \\
\multicolumn{2}{c}{UNSTABLE (Tier 1)} &  &  &  &  &  &  &  &  &  \\
ORB-v2 & $-$ & None & $-$ &  & $-$ & $-$ & $-$ & $-$ & $-$ & $-$ \\
ORB-D3-v2 & $-$ & None & $-$ &  & $-$ & $-$ & $-$ & $-$ & $-$ & $-$ \\
 &  &  &  &  &  &  &  &  &  &  \\
\bottomrule
\end{tabular}
\end{table}

\Needspace{0.45\textheight}
\section*{S2 Angle comparison during MD simulation}\nopagebreak

\begin{figure}[H]
\centering
\includegraphics[width=0.98\textwidth]{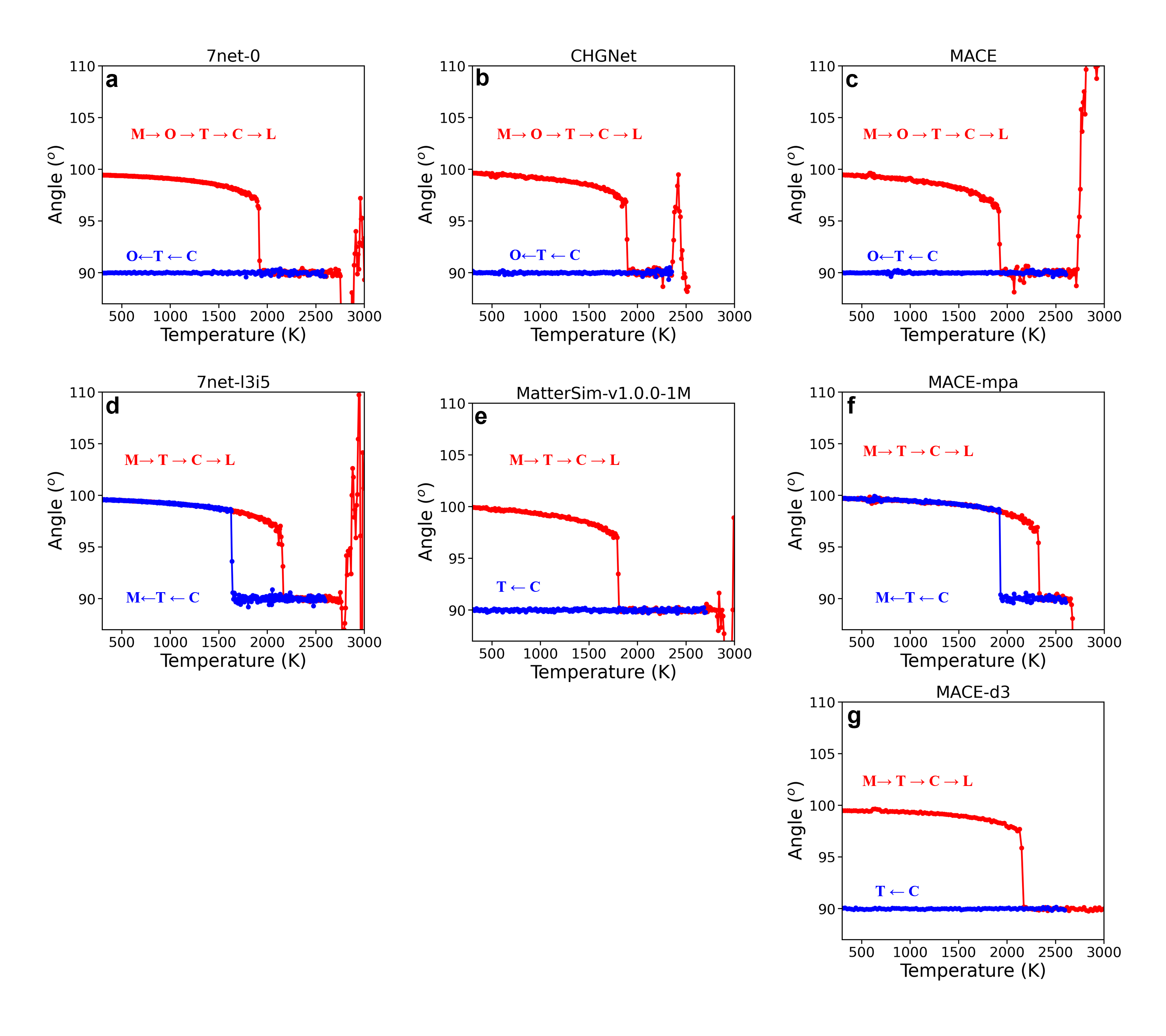}
\caption{\textbf{Evolution of the monoclinic angle $\beta$ during NPT heating and cooling simulations.} Cell angle $\beta$ as a function of temperature for (a) 7net-0, (b) CHGNet, (c) MACE, (d) 7net-l3i5, (e) MatterSim-v1.0.0-1M, (f) MACE-mpa, and (g) MACE-D3. Red curves show heating from the monoclinic phase, and blue curves show cooling from the cubic phase. The top row shows models following the M$\rightarrow$O$\rightarrow$T$\rightarrow$C$\rightarrow$L pathway on heating and O$\leftarrow$T$\leftarrow$C on cooling; the remaining panels show models reproducing M$\rightarrow$T$\rightarrow$C$\rightarrow$L on heating, with cooling behavior indicated in each panel.}
\label{fig:S1}
\end{figure}

\Needspace{0.45\textheight}
\section*{S3 Supercell size effect on phonon dispersion}\nopagebreak

\begin{figure}[H]
\centering
\includegraphics[width=0.70\textwidth]{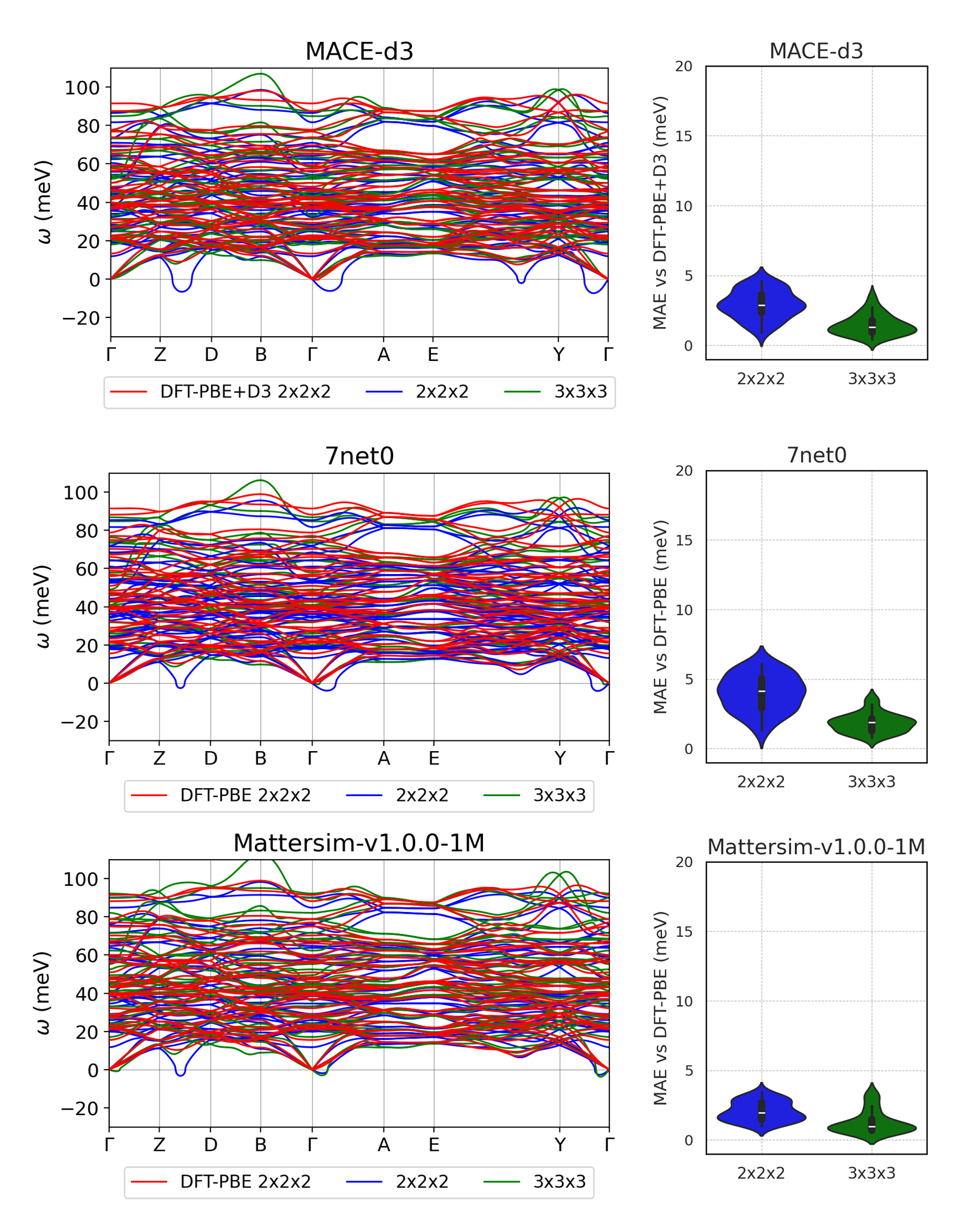}
\caption{\textbf{Supercell-size effect on the monoclinic ZrO}\textbf{\textsubscript{2}} \textbf{phonon dispersion.} Phonon dispersions for the monoclinic phase computed with MACE-D3, 7net0, and MatterSim-v1.0.0-1M with supercell sizes 2x2x2 and 3x3x3, compared against the DFT refences (PBE+D3 for MACE-D3 and plain PBE for 7net0 and MatterSim). Small imaginary acoustic branches near $\Gamma$ are present in 2x2x2 calculations for all uMLFFs. Upon increasing the supercell size to 3x3x3, these near- $\Gamma$ instabilities are removed for MACE-D3 and 7net0, while they persist for MatterSim-v1.0.0-1M. Similarly, soft acoustic modes appearing along the Z-D and E-Y regions for the 2x2x2 supercells are reduced upon supercell enlargement, indicating that they primarily originate from force-constant truncation effects in the smaller supercells for uMLFF.}
\label{fig:S2}
\end{figure}

\Needspace{0.45\textheight}
\section*{S4 Phonon dispersion of all phases of ZrO2 using all models}\nopagebreak

\begin{figure}[H]
\centering
\includegraphics[width=1.00\textwidth]{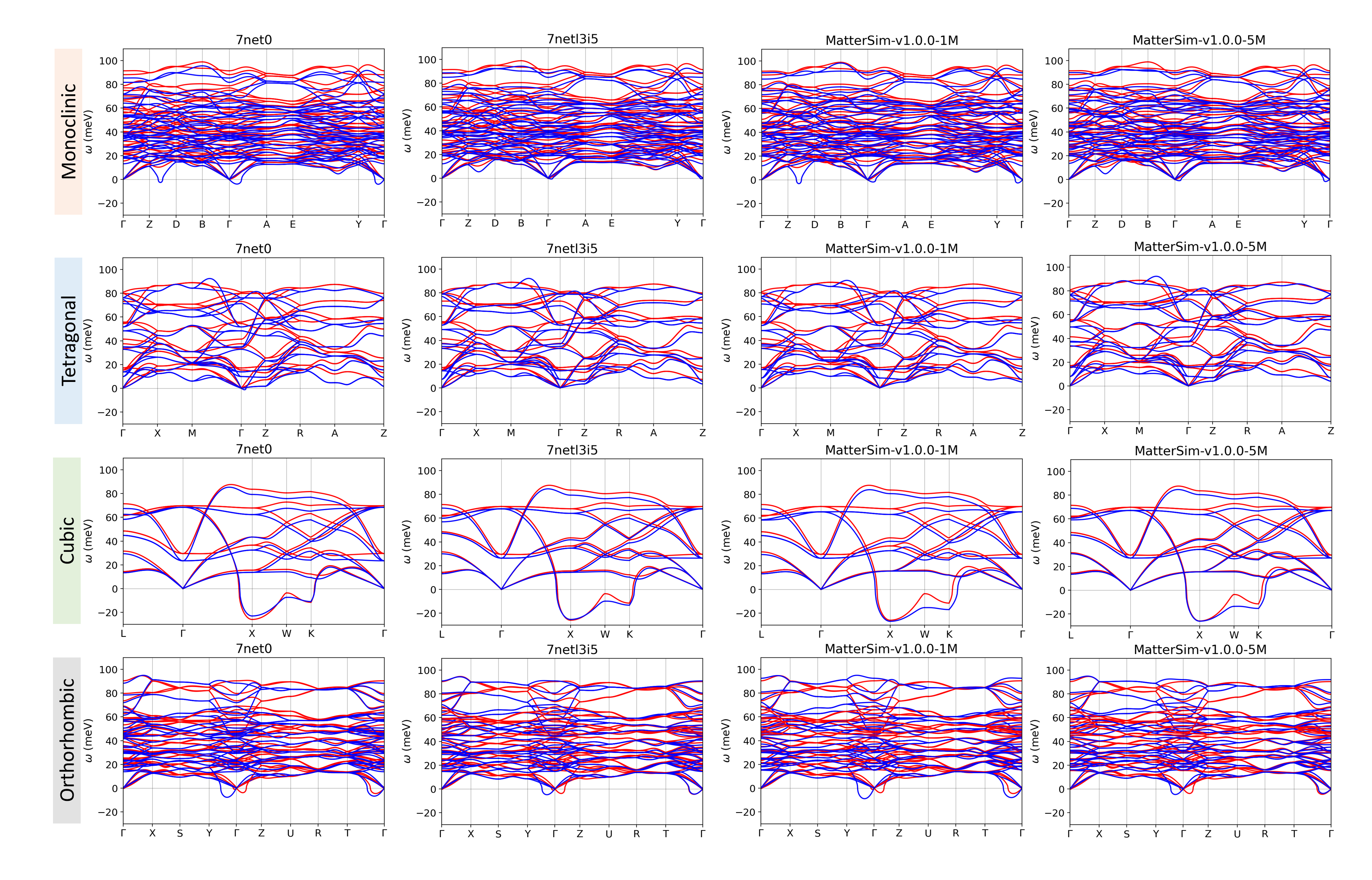}
\caption{\textbf{Phonon dispersions of ZrO}\textbf{\textsubscript{2}} \textbf{polymorphs predicted by uMLFFs and compared with DFT-PBE.} Phonon dispersions for the monoclinic, tetragonal, cubic, and orthorhombic phases (rows) computed with 7net-0, 7net-l3i5, MatterSim-v1.0.0-1M, and MatterSim-v1.0.0-5M (columns). Red curves show DFT-PBE results, and blue curves show uMLFF predictions along the corresponding high-symmetry Brillouin-zone paths.}
\label{fig:S3}
\end{figure}

\begin{figure}[H]
\centering
\includegraphics[width=1.00\textwidth]{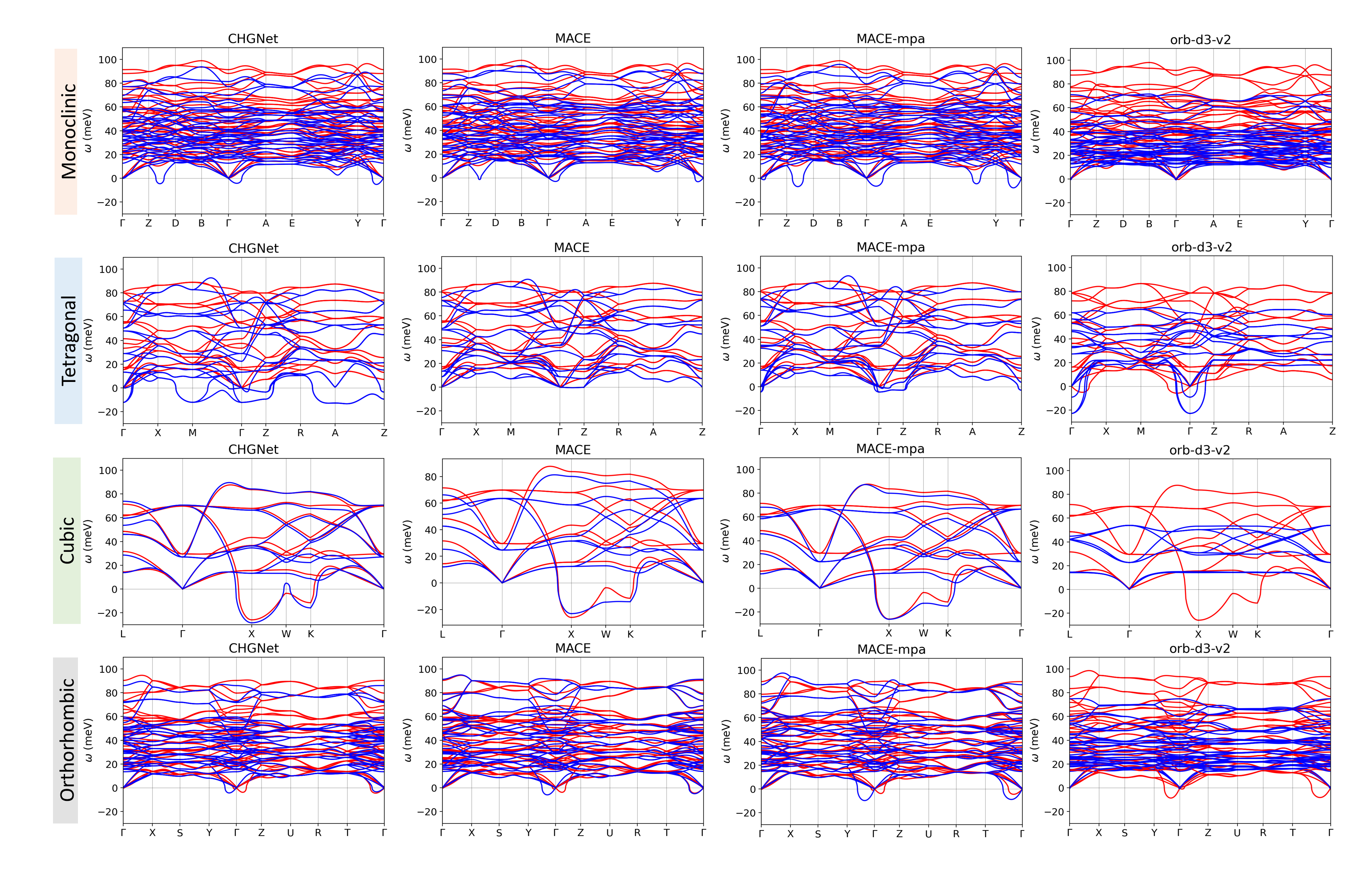}
\caption{\textbf{Phonon dispersions of ZrO}\textbf{\textsubscript{2}} \textbf{polymorphs predicted by uMLFFs and compared with DFT reference calculations.} Phonon dispersions for the monoclinic, tetragonal, cubic, and orthorhombic phases (rows) computed with CHGNet, MACE, MACE-D3, and MACE-mpa (columns). Blue curves show uMLFF predictions, and red curves show the corresponding DFT reference: DFT-PBE for CHGNet, MACE, and MACE-mpa, and DFT-PBE+D3 for MACE-D3.}
\label{fig:S4}
\end{figure}

\section*{S5 Finite temperature corrections}\nopagebreak

Here, we rationalize the emergence of an M$\rightarrow$O transition in terms of the 0 K O--T energy split E\textsubscript{OT} and its deviation from DFT-PBE. We then examine how finite temperature modifies the effective energetic driving forces observed during MD transitions by comparing energy differences extracted from MD trajectories with the corresponding 0 K EOS values.

We focus on 7net-0, 7net-l3i5, and MatterSim-v1.0.0-1M. In the standard Parrinello-Rahman NPT simulations, 7net-0 exhibits an O$\leftarrow$T transition during cooling (see Figure 9), whereas 7net-l3i5 shows an M$\leftarrow$T transition and MatterSim-v1.0.0-1M does not show a transition below T (see Figure 10). However, for both 7net-l3i5 and MatterSim-v1.0.0-1M, the EOS ordering implies that on cooling the system might follow M$\leftarrow$O$\leftarrow$T$\leftarrow$C; however, an O$\leftarrow$T transition was not observed in our standard cooling trajectories. To directly probe the O--T driving force in these models and enable comparison with 7net-0, we therefore performed additional NPT heating simulations initialized from the O phase for 7net-l3i5 and MatterSim-v1.0.0-1M, with fixed cell angles, and confirmed that both models undergo an O$\rightarrow$T transition under these constrained conditions (Figure 14).

During the O--T transition, we obtain E\textsubscript{OT} (T = 2100 K) $\approx$ 35 meV/atom for 7net-0 (Figure 14a). In contrast, for 7net-l3i5 and MatterSim-v1.0.0-1M we find much smaller values, E\textsubscript{OT} $\approx$ 5 meV/atom, at transition temperatures of 1680 K and 1640 K, respectively (Figure 14b,c). Relative to their 0 K EOS values, the finite-temperature contribution inferred from MD reduces E\textsubscript{OT} by 2 meV/atom for 7net-l3i5 and by 4 meV/atom for MatterSim-v1.0.0-1M. By contrast, even at finite temperature, E\textsubscript{OT} for 7net-0 remains nearly twice its 0 K value. This difference is consistent with the observation that 7net-0 favors an O$\leftarrow$T transition during cooling because of the large energy gain, whereas the corresponding driving force is much smaller and further reduced at finite temperature for 7net-l3i5 and MatterSim-v1.0.0-1M.

An analogous comparison can be made for the tetragonal-to-monoclinic transformation in 7net-l3i5, which does exhibit an M$\leftarrow$T transition during cooling. Although not shown in Figure 14, the energy change at the M$\leftarrow$T transition is E\textsubscript{MT} (T = 1635 K) $\approx$ 39 meV/atom, 3 meV lower than E\textsubscript{MT} (T = 0 K) $\approx$ 42 meV/atom. This reduction is consistent with a modest finite-temperature contribution inferred from MD to the relative phase energetics.

\begin{figure}[H]
\centering
\includegraphics[width=1.00\textwidth]{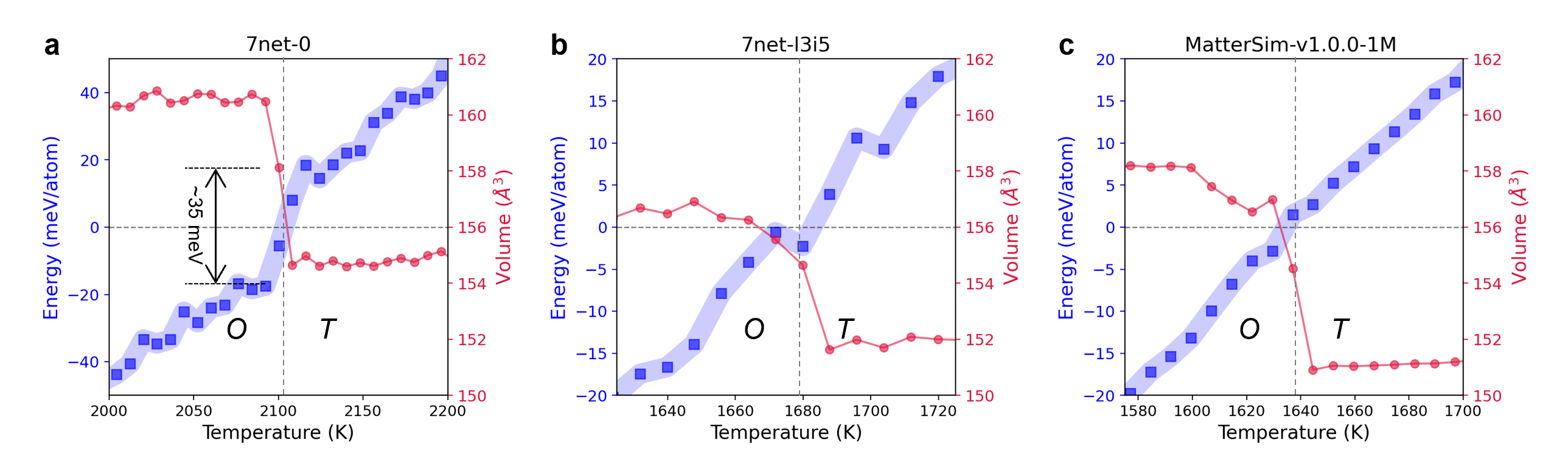}
\caption{\textbf{Volume and energy per atom across the orthorhombic-to-tetragonal transition for (a) 7net-0, (b) 7net-l3i5, and (c) MatterSim-v1.0.0-1M.} The vertical dashed line marks the transition temperature, and phase labels indicate the O and T regions. Blue squares show volume (right axis), and red circles show energy per atom (left axis), referenced to the O-phase energy. For 7net-l3i5 and MatterSim-v1.0.0-1M, trajectories were obtained from NPT heating runs initialized in the O phase with fixed cell angles to induce the O$\rightarrow$T transition.}
\label{fig:S5}
\end{figure}

\Needspace{0.45\textheight}
\section*{S6 Atomic coordinates of the O phase predicted in Section 2.2.1}\nopagebreak

\begin{figure}[H]
\centering
\includegraphics[width=0.35\textwidth]{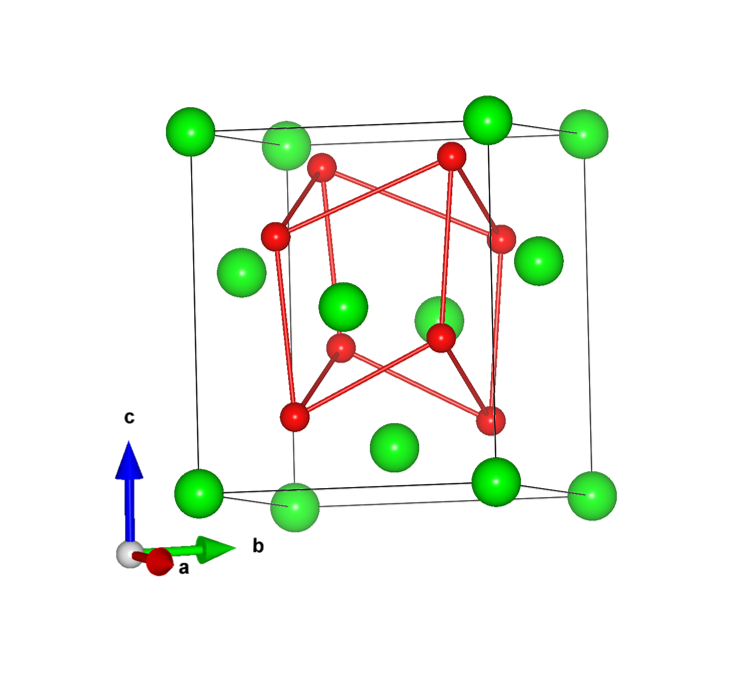}
\caption{\textbf{Orthorhombic ZrO}\textbf{\textsubscript{2}} \textbf{stabilized in uMLFF-driven NPT simulations.} The orthorhombic ZrO\textsubscript{2} phase was observed during NPT heating and cooling simulations using 7net-0, CHGNet, and MACE. Green and red spheres denote Zr and O atoms, respectively.}
\label{fig:S6}
\end{figure}

Below we have the unit-cell details and atomic coordinates of O-phase predicted in Section 2.2.1.

\medskip
{\small
\begin{alltt}
Lattice parameters
 a = 5.30565 \AA{}, b = 5.01550 \AA{}, c = 5.84143 \AA{}, \(\alpha\)=\(\beta\)=\(\gamma\)= 90\textsuperscript{\(\circ\)}
Unit-cell volume = 155.443572 \AA{}\textsuperscript{3}
Atomic-coordinates 
                     x          y          z
  1 Zr  Zr1       1.00000    0.50000    0.50000
  2 Zr  Zr2       0.50000    0.50000    0.13000
  3 Zr  Zr3       1.00000    1.00000    0.00000
  4 Zr  Zr4       0.50000    1.00000    0.63000
  5 O   O1        0.32000    0.27000    0.42000
  6 O   O2        0.17000    0.72000    0.21000
  7 O   O3        0.67000    0.72000    0.42000
  8 O   O4        0.82000    0.27000    0.21000
  9 O   O5        0.17000    0.77000    0.71000
 10 O   O6        0.32000    0.22000    0.92000
 11 O   O7        0.67000    0.77000    0.92000
 12 O   O8        0.82000    0.22000    0.71000
\end{alltt}
}
\medskip

\Needspace{0.45\textheight}
\section*{S7 Atomic coordinates of the O phase from the NIST JARVIS database}\nopagebreak

\begin{figure}[H]
\centering
\includegraphics[width=0.80\textwidth]{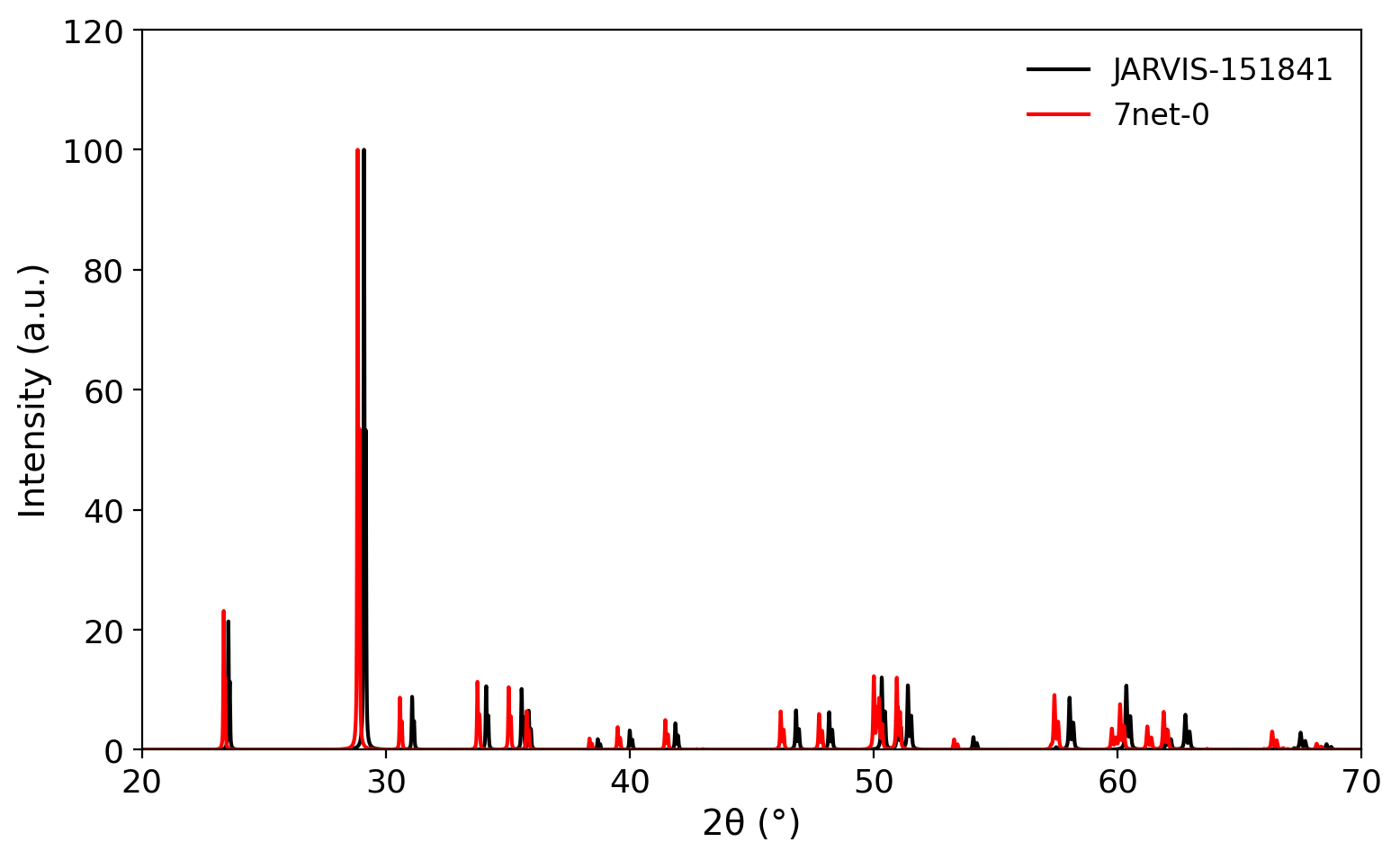}
\caption{\textbf{Powder XRD comparison of JARVIS and uMLFF-stabilized orthorhombic ZrO}\textbf{\textsubscript{2}}\textbf{.} Powder XRD patterns computed with VESTA for the orthorhombic ZrO\textsubscript{2} structure from the JARVIS database (JVASP-151841; black) and the energy-minimized orthorhombic structure from 7net-0 simulations (red), showing that both correspond to the same crystal structure.}
\label{fig:S7}
\end{figure}

Below we have the unit-cell details and atomic coordinates of O-phase ZrO2 \href{https://www.ctcms.nist.gov/~knc6/static/JARVIS-DFT/JVASP-151841.xml}{JARVIS-151841}.

\medskip
{\small
\begin{alltt}
Lattice parameters
a = 5.25187 \AA{}, b = 5.74983 \AA{}, c = 5.00399 \AA{}, \(\alpha\)=\(\beta\)=\(\gamma\)= 90\textsuperscript{\(\circ\)}
Unit-cell volume = 151.107242 \AA{}\textsuperscript{3}
Atomic-coordinates
                     x          y          z
  1 Zr  Zr1       0.00000    0.50000    0.50002
  2 Zr  Zr2       0.50000    0.37478    0.00000
  3 Zr  Zr3       0.00000    0.00000    0.00000
  4 Zr  Zr4       0.50000    0.87478    0.50002
  5 O   O1        0.82274    0.78919    0.27354
  6 O   O2        0.32274    0.08560    0.22648
  7 O   O3        0.82274    0.28919    0.22648
  8 O   O4        0.32274    0.58560    0.27354
  9 O   O5        0.67735    0.58568    0.72658
 10 O   O6        0.17735    0.22650    0.77344
 11 O   O7        0.67735    0.02307    0.77344
 12 O   O8        0.17735    0.78910    0.72658
\end{alltt}
}
\medskip

The powder XRD comparison confirms that the JARVIS structure (JVASP-151841) and the 7net-0-predicted orthorhombic phase correspond to the same crystal structure. The small peak shifts and minor differences in lattice parameters and unit-cell volume arise because the JARVIS entry is a DFT-relaxed structure with slightly different equilibrium lattice constants.

\end{document}